\documentstyle[11pt,arnps]{article}

\def\as{\alpha_{\rm S}}

\def\n{\!\!}
\def\b{\beta}
\def\a{\alpha}
\def\c{\chi   }
\def\d{\delta}

\def\eps{\epsilon}
\def\g{\gamma}

\def\l{\lambda}
\def\L{\Lambda}
\def\m{\mu}
\def\n{\nu}
\def\oa{\omega}

\def\ol{\omega_L}

\def\p{\phi}
\def\pa{\partial}

\def\ra{\rightarrow}

\def\s{\sigma}

\def\ti{\tilde}

\def\({\left(}
\def\){\right)}

\def\citenum#1{{\def\@cite##1##2{##1}\cite{#1}}}
\def\citea#1{\@cite{#1}{}}

\def\d{\mathop{\rm d}}

\def\Im#1{\mathop{\rm Im}\{#1\}}
\def\Re#1{\mathop{\rm Re}\{#1\}}

\def\beq{\begin{equation}}
\def\eeq{\end{equation}}
\def\bea{\begin{eqnarray}}
\def\eea{\end{eqnarray}}
\def\eq#1{{eq.~(\ref{#1})}}
\def\eqs#1#2{{eqs.~(\ref{#1})--(\ref{#2})}}
\def\bbbz{{\mathchoice {\hbox{$\sf\textstyle Z\kern-0.4em Z$}}
{\hbox{$\sf\textstyle Z\kern-0.4em Z$}}
{\hbox{$\sf\scriptstyle Z\kern-0.3em Z$}}
{\hbox{$\sf\scriptscriptstyle Z\kern-0.2em Z$}}}}
\def\npb#1#2#3{    {\it Nucl. Phys. }{\bf B#1} #2 #3}
\def\plb#1#2#3{    {\it Phys. Lett. }{\bf B#1} #2 #3}
\def\prd#1#2#3{    {\it Phys. Rev. }{\bf D#1} #2 #3}
\def\prep#1#2#3{   {\it Phys. Rep. }{\bf #1} #2 #3}
\def\prl#1#2#3{    {\it Phys. Rev. Lett. }{\bf #1} #2 #3}

\def\rmp#1#2#3{    {\it Rev. Mod. Phys. }{\bf #1} #2 #3}
\def\zpc#1#2#3{    {\it Z. Phys. }{\bf C#1} #2 #3}

\def\sjnp#1#2#3{   {\it Sov. J. Nucl. Phys. }{\bf #1} #2 #3}

\def\ppjetp#1#2#3{ {\it Sov. Phys. JETP }{\bf #1} #2 #3}
\def\ppjetpl#1#2#3{{\it JETP Lett. }{\bf #1} #2 #3}
\def\zetf#1#2#3{   {\it Zh. ETF }{\bf #1}#2 #3}

\def\jpg#1{        {\it J. Phys}. {\bf G#1}}
\relax

\begin{document}
\begin{flushright}
\begin{Large}
 TAUP 2221-94 \\
 CBPF NF 001/95\\
December 1994\\
\end{Large}
\end{flushright}
\vskip 2 true cm
\begin{center}
\Huge {\bf RENORMALONS at LOW x }

 \end{center}
\vskip 4pt

\begin{center}

\begin{Large}
EUGENE LEVIN $^{\bullet}$ $^{\dagger}$
\footnotetext{$^{\bullet}$  On leave of
Theory Department,
St. Petersburg Nuclear Physics Institute,\\
 188350, St. Petersburg, Gatchina, RUSSIA}
\footnotetext{ $^{\dagger}$ Email:\,\, levin@lafex.cbpf.br;\,\,
levin@ccsg.tau.ac.il}
 \end{Large}
\par \vskip 0.2 true cm
\noindent
{\sl Mortimer and Raymond Sackler Institute of Advanced Studies\\
School of Physics and Astronomy, Tel Aviv University\\
Ramat Aviv, 69978, ISRAEL}\\
and\\
{\sl LAFEX, Centro Brasileiro de Pesquisas F\'\i sicas / CNPq
\\Rua Dr. Xavier Sigaud 150, 22290 - 180, Rio de Janiero, RJ, BRASIL}
\end{center}
\par \vskip .5in

\begin{abstract}
\par \vskip .3in \noindent
The role of infrared and ultraviolet renormalons are discussed in context
of leading log(1/x) approximation of perturbative QCD. We generalize the BFKL
equation for the case of running coupling QCD constant and show that the
uncertainties
 related to the contribution of infrared renormalons turn out  to be smaller
than  the
shadowing correction to the total cross section. The contribution of
infrared and ultraviolet renormalons to the BFKL equation are studied,  and
the solution of the BFKL equation with running coupling constant is
discussed.

 \end{abstract}
\par \vskip 2in \noindent

\pagebreak
\section{Introduction.}
In this paper we study two problems:

1. How to include the running coupling QCD constant in the BFKL equation
\cite{BFKL};

2. The  contributions  to the solution of the BFKL equation which  originate
from infrared and ultraviolet renormalons.

To demonstrate why these two problems are important for the low x physics,
we  review the main points of our strategy in perturbative
QCD approach.
 We  start with perturbative QCD ( pQCD )  in the kinematic  region
where the
 parton  density and coupling QCD constant ($\alpha_s$) are small.
Each physical observable (i.e.  gluon structure function) can  be written in
pQCD as following series:
\beq \label{PERSER}
xG(x,Q^2)\,\,=\,\,\Sigma_{n = 0} \,C_n ( \alpha_s)^n\cdot(L^n + a_{n - 1}
L^{n - 1} ... a_0)\
,\,.
\eeq
This perturbative series  has  two big problems:

{\bf 1.} The natural small parameter $\alpha_s$ is compensated by large log
(L).
The value of L depends on the process and kinematic region. For example in
deeply inelastic scattering (DIS):
$$ L \,\,=\,\,\log \frac{ Q^2}{Q^2_0} \,\,\,\,\,\,at\,\,\,\, Q^2 \gg
Q^2_0\,\,\,\,but \,\,\,\,\,\,x\,\sim\,1 $$
$$
\,\,L\,\,=\,\,\log(1/x)\,\,\,\,\,\,\,at\,\,\,\,\,\, Q^2\,\sim\,Q^2_0
\,\,\,\,\,\,and\,\,\,\,x\,\rightarrow\,0 $$
$$
\,\,\,\,\,\,\,\,\,\,\,L\,\,=\,\log Q^2\,\cdot\,\log(1/x)\,\,at\,\,\,\, Q^2\,\gg
Q^2_0 \,\,\,\,\,\,and \,\,\,\,x\,\rightarrow \,0
$$
$$
\,\,L\,\,=\,\,\log(1 - x)\,\, \,\, at\,\,\,\, Q^2\,\sim\,\,Q^2_0
\,\,\,\,\,\,and\,\,\,\,\,\, x\, \rightarrow\,1
$$
Of course it is not the full list of scales. The only that we would like to
 emphasize that $L$ depends on the kinematic region. Thus to calculate
 $xG(x,Q^2)$ one cannot calculate only the Born Approximation, but has to
calculate a  huge number of Feynman diagrams.

{\bf 2.}
$
\,\,\,\,\,C_N\,\,\rightarrow\,\,n! \,\,\,\,\,at\,\,\,\, n\,\,\gg\,\,1\,
\,\cite{N!}. $
Means that we are dealing with an  asymptotic series and we do not know the
 general rules of what to do with such series. There is only one rule,
namely to
 find the analytic function which has the same perturbative series. Sometimes
but very rarely we can find such analytic function. For  this case,  this
is the
 exact solution of our  problem. However,  the general
approach has been developed
based on Leading Log Approximation (LLA). The idea is simple. Let us find
the analytic function that sums the series:
\beq \label{LLA}
xG(x,Q^2)_{LLA}\,\,=\,\,\Sigma_{n=0} C_n\, (\alpha_s \cdot L )^n\,\,.
\eeq
Usually we can write the equation for function $xG(x,Q^2)_{LLA}$. The most
famous one, the GLAP evolution equation \cite{GLAP}, sums eq.(2) if $L$=
$\log( Q^2/Q^2_0)$.  However, it turns out that for the region of small $x$
the most important  is so called the  BFKL \cite{BFKL} equation.
 It  gives the answer for
eq.(2) in the
case when $L$=$\log(1/x)$.

 Using the solution of the LLA equation we build
the ratio:
\beq
R(x,Q^2)\,=\,\frac{xG(x,Q^2)}{xG(x,Q^2)_{LLA}}\,\,=\,\,\Sigma_{n=0} r^n\,=
\Sigma_{n=0} c_n\cdot(L^{n-1} + a_{n-1}L^{n - 2} + ... a_0)\,\,.
\eeq
This ratio is also asymptotic series, however here  we calculate
this series term by term. Our hope is that the value of the next term will be
 smaller then the previous one $( \frac{r_{n} }{r_{n - 1}} \ll 1 )$ for
 sufficiently large $n$. However,  we know that at some value of $n=N$,
$\frac{r_{N}}{r_{N-1}}\,\sim \,1$. The only thing  that we can say about such
situation
 in  general, is  that our calculation has intrinsic theoretical accuracy,
and the
 result of calculation should be presented in the form
\beq
R(x,Q^2)\,\,=\,\,\Sigma^{n = N-1}_{n=0} r_{n} \,\pm r_{N}\,\,.
\eeq
  The size of $N$ depends mostly on how well we chose the LLA,  and how
well we established the value of scale $L$ in the process of interest.

In the region of small x the natural value of the scale $L$ is $ L = \ln
(1/x)$ and the series of eq.(~\ref{LLA}) can be summed by the BFKL equation.
 We have reached substantial understanding of
the main properties of this equation during the last two decades ( see
ref.\cite{LIPUN}). The last major advance
in such understanding was made by A.Mueller \cite{MU} who introduced the
correct degrees of freedom - colour dipoles, in terms of which the BFKL
equation got the partonic-like probabilistic interpretation ( see also
related papers \cite{REMU}). However one problem with the BFKL equation is
still open,namely we do not know how to include the running QCD coupling
constant in it.

 There are two aspects of the problem. The first one is of
the fundamental importance. Indeed, only the BFKL equation with running
coupling
constant can turn into the GLAP equation at large value of virtualities $Q^2$.
We  discuss this issue in the next section. The second aspect is
more practical one. The first attempts to take into account the running
coupling constant  ( $ \as$) in the BFKL equation \cite{FIRST} show that the
behaviour
of the solution at low x which is power-like ( $ x G (x,Q^2) \propto
(\frac{1}{x})^{\,\omega_0}$) becomes much slower ( the value of $ \omega_0$
is significantly lower )  than for the original version of the BFKL equation
with fixed  $\as$. In section 2 we will generalize the BFKL equation for  the
case of running $\as$.

The question arises: can we guarantee the accuracy of the BFKL equation with
running $\as$ knowing that the calculation even the correction of the
order of
$\as$ to the BFKL equation is a  very difficult task,  which to date  has not
been
performed even by the great experts in the field \cite{LIFA}. To answer this
question, we have to discuss the second problem in the perturbative series,
 namely the $n!$ rise of coefficients $ C_n$ ( see eq.~\ref{PERSER} ).

By now we have known three sources of the $n!$ behaviour of $C_n$: infrared
(IR)
and ultraviolet ( UV) renormalons which are intimately related to the running
$\as$
and instanton contribution. In this paper we concentrate our efforts on the
first two, because the instantons at high energy give negligible
contribution
 ( see refs. \cite{INSTANTON} for relevant discussion). The $n!$
behaviour of $C_n$ from IR and UV renormalons  originate from running $\as$
and real parameter which governs the inclusion of $\as$ is $\as^n\,n!$.

In section 3,  we study the first correction to the BFKL equation due to
running $\as$ in detail,  and show three  principle results. The first one is
that we can absorb all uncertainties related to the contribution of the IR
renormalons,
which have basically nonperturbative origin,  to the shadowing
correction (SC) to the total cross section.
 It should be stressed that those SC can be expressed through the
correlation length between two gluons, which cannot be calculated in the
framework of perturbative QCD. The second result is the fact that we can
guarantee the accuracy of the BFKL equation with running $\as$ summing all
terms of the order $\as^n n!$ in the series of eq. (~\ref{PERSER}).
The third result is the nonperturbative contribution of the order of
$\sqrt{\frac{\L^2}{Q^2}}$ from ultraviolet renormalons which does not
appear in the Wilson Operator Expansion.

In section 4  we are going to discuss the solution to the BFKL equation with
running $\as$. We start with the numerical estimates of the first correction
to the value of $\omega_0$ due to running $\as$. We formulate our numerical
accuracy in the attempts to solve the BFKL equation with running $\as$ and
find the Green function for  it. We show that running $\as$ leads to
 a  smaller value of $\omega_0$ ($ x G(x, Q^2) \ra
(\frac{1}{x})^{\omega_0}$, than the  original  BFKL equation with fixed $\as$.

\section{ The BFKL equation with running QCD coupling constant.}
In this section we  discuss the BFKL equation \cite{BFKL} with
running $\as$, but  start with the original version of this equation with
fixed $\as$, both for the sake of completeness of the presentation and to
clarify the main steps in the derivation of the BFKL equation. It should be
mentioned that we will follow the original derivation of the BFKL equation in
the momentum representation since we found  it more economic to include
the running $\as$ than the Mueller approach \cite{MU}. We firmly believe
that it could also be duplicated  in dipole picture, but leave this job
for further publication.
\subsection{ The BFKL equation in the lowest order of $\as$.}
 The BFKL equation  was derived in so called Leading Log (1/x)
Approximation, in
which we would like to keep the contribution of the order of $( \as log
(1/x))^n$ and neglect all other contributions, even of the order of $\as\,\,
log (Q^2/Q^2_0)$. So the set of parameters in  LL(1/x)A is obvious:
$$
\as\,\, log \frac{1}{x} \,\,\sim\,\,1\,\,;
$$
\beq
\as\,\, log \frac{Q^2}{Q^2_0} \,\,< \,\,1\,\,;
\eeq
$$
\as\,\,\,\ll\,\,1\,\,;
$$
Let us consider the simplest process: the quark - quark scattering at high
energy at zero  momentum transfer . All problems of infrared divergency in
such a process is irrelevant since they are canceled in the scattering of two
colourless hadrons (see, for example, ref. \cite{MU} for details).

In the Born Approximation the only diagram of Fig. 2.1  contributes to the
imaginary part of the scattering amplitude (A). It is easy to understand
that the result of calculation of this diagram gives:
\beq \label{BA}
 2 \Im{A^{BA}(s,t = 0 )}\,\,=\,\,\int \frac{ d^2 k_t}{ ( 2 \pi)^2} |\,\, M ( 2
\ra 2;\, \as\,\, |\,\, s, t = - k^2_t)\,\, |^2
\,\,=\,\,s \frac{C^2_2 \as^2}{ N^2 - 1} \,\int\,\frac{ d^2 k_t}{ k^4_t}\,\,,
\eeq
where $ M (2 \ra 2;\, \as\, |\, s, t = - k^2_t ) $ denotes the amplitude in
the lowest order of $\as$ for
quark - quark scattering at transfer momentum $ t= - k^2_t $ through one gluon
exchange ( see Fig.2.1 ),
N is the number of colours and $C_2 \,=\,\frac{ N^2 - 1}{ 2 N} $. One can
recognize the Low - Nussinov mechanism \cite{LONU} of high energy
interaction in this simple example.

In the next order we have to consider a larger  number of the diagrams, but
 we can write down the answer in the following general form:
\beq \label{NEBA}
2 \Im{A^{NBA} ( s,t=0)} \,\,=\,\,
\eeq
$$
\int ( 2 \pi )^4 \delta^{(4)}( {\bf p_1 + p_2 -
p'_1 - p'_2 - q } ) |\,\, M ( 2 \ra 3; g^3)\,\,|^2\,\, \Pi^{i = 3}_{i = 1}
\frac{d^3 p'_i}{ ( 2 \pi )^3  2 E'_i}  \,\,+$$
$$
 \int ( 2 \pi )^4 \delta^{(4)} ( {\bf p_1 +
p_1 - p'_1 - p'_2})\cdot 2\, \Re{
 M ( 2 \ra 2 ; \as^2 ) \cdot M^{*} (2 \ra 2 ; \as)}
 \Pi^{ i = 2}_{i = 1}\,\, \frac{ d^3 p'_i}{ ( 2 \pi)^3 2 E'_i}\,\,,
$$
where $g$ is coupling constant of QCD ( $\as = \frac{g^2}{4 \pi}$);
$ M( 2 \ra 3; g^3)$ is the amplitude for the production of the extra gluon
in the Born Approximation, and  is given by the set of Feynman diagram in
Fig.2.2  while $ M ( 2 \ra 2, \as^2 ) $ is the amplitude of the elastic
scattering in the next to leading Born Approximation ( see Fig. 2.3) at the
momentum transfer $k_t$.

Two terms in \eq{NEBA} have different physical meaning: the first one
describes the emission of the additional gluon in the final state of our
reaction,  while the second term is the virtual correction to the Born
Approximation due to the emission of the additional gluon. It corresponds to
the same two particle final state, and describes the fact that due to
emission the probability to detect this final state becomes smaller ( we
will see later that the sign of the second term is negative).

In the both terms of \eq{NEBA} we can integrate over ${ \bf \vec{p_3}}'$ as
well as
over the longitudinal component of ${\bf {\vec  p'_1}}$ ($p'_{1L}$).
Finally, we  rewrite the phase space in the following way:
$$
\int ( 2 \pi )^4 \delta^{(4)}( {\bf p_1 + p_2 -
p'_1 - p'_2 - q }\,\, \Pi^{i = 3}_{i = 1}\frac{d^3 p'_i}{ ( 2 \pi )^3  2
E'_i}\,\,=\,\,\frac{1}{ 4 \pi}\cdot\frac{1}{s} \cdot \int^{1}_{
x_{min} = \frac{m^2}{s}}
\frac{d x'_3}{x'_3} \int \frac{ d^2 p'_{1t} d^2 p'_{3t}}{ ( 2\pi )^4}\,\,;
$$
\beq \label{PS}
\int ( 2 \pi )^4 \delta^{(4)} ( {\bf p_1 +
p_1 - p'_1 - p'_2})\cdot\Pi^{ i = 2}_{i = 1}\,\,
 \frac{ d^3 p'_i}{ ( 2 \pi)^3 2 E'_i}\,\,=\,\, \frac{1}{ s} \cdot \int
\frac{d^2 p'_{1t}}{ ( 2 \pi)^2}\,\,;
\eeq
where $x$ is the fraction of the longitudinal momentum carried by particle.
the value of $x_{min}$ depends on the  reaction. For example in the deeply
inelastic scattering $x_{min}\,=\,x_{B}\,=\,\frac{|Q^2|}{s}$. In the case of
the quark scattering $x_{min} \,=\,\frac{m^2_t}{s}$ where $m_t$ is the
transverse mass of produced quark.

 From \eq{PS} one can see the origin of the $log (1/x_{min})$ contribution:
it stems from the phase space integration, if $ M ( 2\,\ra\,3 )$ does not
go
to zero at $ x_3 \,\ra \, 0 $. To sum the diagrams of Fig. 2.2  in this
limit we can use two tricks. The first one is  for each $t$-
channel gluon one can rewrite the numerator of the gluon propagator at high
energy in the following way:
\beq \label{GLUON}
g_{\m \n}\,\,=\,\,\frac{ p_{1 \m} p_{2 \n} \,\,+\,\,p_{2 \m} p_{1 \n}}{
p_1\cdot p_2}\,\,+\,\, O
(\,\, \frac{m^2}{ s}\,\,)
\eeq
The second trick is based on the gauge invariance of the QCD. We look on
the subset of the diagrams of Fig. 2.2  pictured in Fig. 2.4  as the
amplitude of the interaction of gluon $ k$ with the quark $p_2$ ( see
Fig.2.4 ).  Since all particles in amplitude $ M_{\n}$ except gluon $k$
are on the mass shell, the gauge invariance leads to the relationship:
\beq \label{GAUGETRICK}
k_{\n} M_{\n} \,\,=\,\,0\,\,.
\eeq
Using Sudakov variables \cite{SUDAKOV} we can expand  vector $k$ as
$$
k_{\n}\,\,=\,\,\a_{k} \,p_{1\n} \,\,+\,\,\b_{k}\,p_{2\m} \,\,+\,\,k_{t \m}
$$
and  rewrite \eq{GAUGETRICK} in the form:
\beq \label{GATR}
\(\,\,\a_{k} \,p_{1\n} \,\,+\,\,\b_{k}\,p_{2\m} \,\,+\,\,k_{t
\m}\,\,\)\,\,M_{\n}\,\,=\,\,0\,\,.
\eeq
 We  note that all particle inside $ M_{\n}$ have a large component of
their momentum on $p_2$. It means that we can neglect the projection of
vector $M_{\n}$ on $p_1$,  or in other words
$$
M_{\n}\,\,=\,\, M^{(1)} p_{1 \n} \,\,+\,\,M^{(2)} p_{2 \n}
\,\,+\,\,M^{(t)}_{\n}
$$
 and $ M^{(1)}\,\ll\,M^{(2)} $.
 Thus we can conclude from \eq{GATR} that
\beq \label{TRPO}
 p_{1 \n} M_{\n}\,\,=\,\,-\,\frac{k_{t \n} M_{\n}}{\a_k}\,\,.
\eeq
 Using both tricks of \eq{GLUON} and \eq{TRPO} one can easily see that only
diagram of Fig. 2.4 ( 1 ) contributes in LL(log(1/x)A. Indeed, let us
consider for example the diagram of Fig. 2.4 ( 2 ), the dominator of the quark
propagator $( p_2 + k )^2 $ is equal to
$$
( p_2\,+\,k )^2\,\,=\,\,\a_k s \,\,+\,\,k^2_t\,\,=\,\,-\,\frac{p'^2_{3t}}{x_3}
\,+\,k^2_t\,\,\sim\,\,-\,\frac{p'^2_{3t}}{x_3}\,\,.
$$
Since due to \eq{TRPO} the polarization of gluon $k$ is transverse we cannot
compensate the smallness this diagram at $x_3 \,\ra\,0$. Using the same
tricks with the upper parts of the diagrams of Fig. 2.2 we arrive at the
conclusion that that the set of the diagrams of Fig. 2.2 degenerates into
one diagram of Fig. 2.5 with specific vertex for gluon emission:
\beq \label{VERTEX}
\Gamma_{\sigma}\,\,=\,\, i\,g\,f_{a b c}\cdot \frac{2 \, k_{t \m}\,k'_{t \n}}
{ \a_k \b_{k'} s}\, \gamma_{\m \sigma \n}\,\,,
\eeq
 where $\gamma_{\m \sigma \n}$ is  given by the usual Feynman rules for QCD.

Substituting  \eq{VERTEX} into the first term of \eq{NEBA} we  get the
contribution of the emission of one additional gluon to the next to Born
Approximation in the form:
\beq \label{EMINBA}
\Im{A^{NBA}_{emission}}\,\,\,\,=\,\,s\,\frac{\as^2\,C^2_2}{ N^2\,-\,1}
\int^y_0 d y' \,\int\,\frac{d^2 k_t}{\pi\, k^4}\,\cdot \frac{ N \as}{\pi}
\cdot  K_{emission} (
k_t, k'_t )\cdot \,\frac{d^2 k'_t}{\pi\, k'^4} \,\,,
\eeq
where $ y'\,=\,log (1/x_3)$, $y \,=\,log (1 / x_{min}$  and
 the kernel $K (k_t, k'_t )$ is equal to
\beq \label{KERNEM}
K_{emission} (k_t, k'_t)\,\,=\,\,\frac{ k^2_t\, k'^2_t}{
(\,k_t\,\,-\,\,k'_t\,)^2} \eeq.

To calculate the virtual correction in the next to Born Approximation
 ($ M (\,2\,\ra\,2\,;\as^2\,)$ )  we have
to estimate the contribution of  the set of diagrams of Fig. 2.3.
The log (1/x) contribution is hidden in the real part of the amplitude
$ M (\,2\,\ra\,2\,;\as^2\,)$, and easiest way to extract this log is to
use the dispersion relation:
\beq \label{DISRE}
M(\,2\,\ra\,2\,;\,\as^2\,)\,\,=\,\,\frac{1}{\pi} \cdot \{\,\,\int
\frac{\Im{M(\,2\,\ra\,2\,;\,\as^2\,)}_s}{ s'\,\,-\,\,s} \,d\,s' \,\,+\,\,
\frac{\Im{M(\,2\,\ra\,2\,;\,\as^2\,)}_u}{ u'\,\,-\,\,u} \,d\,u'\,\,\}\,\,.
\eeq
We calculate  $\Im{M(\,2\,\ra\,2\,;\,\as^2\,)}_s$ and
$ \Im{M(\,2\,\ra\,2\,;\,\as^2\,)}_u$ using unitarity ( see \eq{BA}):
\beq \label{UNITARITY}
 \Im{M(\,2\,\ra\,2\,;\,\as^2\,\,| \,t\,=\, - k^2_t\,)}_s\,\, =
\eeq
$$
\,\,\int \frac{ d^2 k'_t}{ ( 2 \pi)^2} |\,\,
\Re{M ( 2 \ra 2;\, \as\,\, |\,\, s, t = - k'^2_t)\,\,
M ( 2 \ra 2;\, \as\,\, |\,\, s, t = - ( k_t - k'_t )^2\,)}\,\,.
$$
The difference between \eq{UNITARITY} and \eq{BA} is that the Born amplitude
for one gluon exchange enters these two equations at different values of
momentum  transferred $t$. The explicit calculations give:
\beq \label{RENBA}
 \Im{M(\,2\,\ra\,2\,;\,\as^2\,\,| \,t\,=\, - k^2_t\,)}_s \,\,=
\,\,s\,\cdot\,C_s\,\pi\,\Sigma ( k^2_t );
\eeq
$$
\Im{M(\,2\,\ra\,2\,;\,\as^2\,\,| \,t\,=\, - k^2_t\,)}_u \,\,=
\,\,s\,\cdot\,C_u\,\pi \,\Sigma ( k^2_t );
$$
$$
\Sigma ( k^2_t ) \,\,=\,\, \frac{4 \as }{\pi^2}\,\int \frac{d^2 k'_t}{ (
\,k_t - k'_t \,)^2\,k'^2_t}\,\,;
$$
Using the dispersion relation of \eq{DISRE} one can reconstruct the real part
of the amplitude and the answer is
\beq \label{RE}
\Re{M(\,2\,\ra\,2\,;\,\as^2\,\,| \,t\,=\, -
k^2_t\,)}\,\,=\,\, s\, (\, C_u \,-\, C_s\,)\cdot \Sigma ( k^2_t ) \cdot log s
\eeq
The colour coefficients have a very famous relation between them ( see Fig.
2.6 ) which gives, for the difference of the colour coefficients in \eq{RE}
the same colour structure as for the diagram of Fig. 2.3 ( 3 ). Thus
$\Re{M(\,2\,\ra\,2\,;\,\as^2\,\,| \,t\,=\, - k^2_t\,)}$
 has the same colour structure as one gluon exchange in the Born Approximation.
This fact makes it  possible to rewrite the second term in \eq{NEBA} as the
correction to the gluon trajectory, e.g. instead of gluon with propagator
$\frac{1}{k^2 }\cdot s $ we can introduce the new propagator
\beq \label{REGGE}
\frac{1}{k^2}\cdot s^{\a^{G}( k^2)}
\eeq
$$
\a^{G}(k^2)\,\,=\,\, 1\,\,-\,\,\frac{\as N}{ \pi^2}\cdot \int \frac{ k^2_t
d^2 k'_t}{ (\,k_t\,-\,k'_t\,)^2 \, k'^2_t}\,\,=\,\,1\,\,-\,\,\frac{\as N}{
 2 \pi^2}\cdot \int \frac{ k^2_t
d^2 k'_t}{[\, (\,k_t\,-\,k'_t\,)^2 \,+\, k'^2_t\,] k'^2_t}
$$
The answer for the second term in \eq{NEBA} can be written in the form
\beq \label{VIRTUAL}
\Im{ A^{NBA}_{virtual}}\,\,=\,\,s\,\frac{\as^2 C^2_2}{ N^2 \,-\,1}\,
\int  \(\, \a^{G} ( k^2 )\,\,-\,\,1\,\)\cdot 2 \cdot \frac{ d^2 k_t}{
k^4_t}\,\,.
\eeq
We can get the full answer for the amplitude in the next to the Born
Approximation ($\as^3$ by summing \eq{EMINBA} and \eq{VIRTUAL} and it can be
written in the form:
\beq \label{FINNBA}
\Im{A^{NBA} ( s, t = 0 )} \,\,=\,\,s\,\frac{\as^2\,C^2_2}{
N^2\,-\,1}\,\,\int^y_0 d y' \int \,\frac{d^2 k_t}{ k^2_t}\cdot\frac{ \as
N}{\pi^2} \cdot K ( k_t, k'_t ) \cdot \frac{d^2 k'_t}{ k'^2_t}
\eeq
where $ y \,=\,log (1/x_{min}$ and
\beq \label{LIKER}
K ( k_t, k'_t ) \cdot \frac{1}{ k'^2_t} \,\,=\,\,
\frac{1}{ (\,k_t\,-\,k'_t\,)^2}\cdot\frac{1}{k'^2_t}\,\,-
\,\,\frac{k^2_t}{(\,k_t\,-\,k'_t\,)^2\,[\, (\, k_t\,-\,k'_t\,)^2\,+\,k'^2_t\,]}
\cdot\frac{1}{k^2_t}\,\,.
\eeq
Using \eqs{FINNBA}{LIKER} we can introduce function $\p ( k^2_t)$ and rewrite
the
total cross section for quark - quark scattering in the form:
\beq \label{QUARKX}
\s_{qq}\,\,=\,\,\frac{\as C_2}{ N^2\,-\,1}\cdot \int \p ( y, k^2_t) \cdot
\frac{ d k^2_t}{ k^2_t}\,\,.
\eeq
For $\p$ \eq{FINNBA} gives the equation:
\beq \label{LIEQNBA}
\p^{(2)} ( y, k^2_t )\,\,=\,\,\frac{ \as N}{ \pi^2} \cdot \int^y_0 d y' \int
d^2 k'_t
K (k_t, k'_t) \,\,\p^{(1)} ( y', k'^2_t )\,\,.
\eeq
where
\beq \label{LIKERNEL}
K ( k_t, k'_t ) \p ( k^2_t ) \,\,=\,\,
\frac{1}{ (\,k_t\,-\,k'_t\,)^2}\cdot \p ( k'^2_t ) \,\,-
\,\,\frac{k^2_t}{(\,k_t\,-\,k'_t\,)^2\,[\, (\, k_t\,-\,k'_t\,)^2\,+\,k'^2_t\,]}
\cdot \p ( k^2_t )\,\,.
\eeq
and
\beq \label{INCONG}
\p^{(1)}\,\,=\,\, \frac{\as C_2}{ k^2_t}\,\,.
\eeq
\subsection{The main property of the BFKL equation.}
 From the simplest calculation in $\as^3$ order one can guess the BFKL equation
\cite{BFKL} which looks as follows:
\beq \label{BFKL}
\frac{d \p (\, y\,=\,log (1/x ), k^2_t\, )}{ d y} \,\,=\,\,\frac{\as
N}{\pi^2} \cdot \int \, K (\,k_t, k'_t\,)\,\p (\,y, k'^2_t\,)\,d^2 k'_t\,\,,
\eeq
where kernel $ K (k_t, k'_t) $ is defined by \eq{LIKERNEL}.
This equation sums the $ ( \as log (1/x ) )^n $ contributions and has
"ladder" - like structure   ( see Fig. 2.7 ). However, such "ladder"
diagrams are only an effective representation of the whole huge set of the
Feynman diagrams, as explained in the simplest example of
the previous subsection. The first part of the kernel $ K (k_t, k'_t )$
describes the emission of new gluon, but with the vertex which differs from
the vertex in the Feynman diagram, while the second one is related to the
reggeization of all t-channel gluons in the "ladder".

 The solution of the BFKL equation has been given in ref. \cite{BFKL} and we
we would like to recall some main properties of this solution.

\subsubsection{Eigenfunctions of the BFKL equation.}
The eigenfunction of the kernel $K(k_t,k'_t)$ is $\p_f\,\,=\,\,(k^2_t)^{f
- 1}$. Indeed after sufficiently long algebra we can see that
\beq \label{EIGENVALUE}
\frac{1}{\pi} \int d^2 k'_t  K (k_t,k'_t)\,\, \,\p_f ( k'^2_t )\,\,=\,\,\c
( f) \,\,\,\p_f ( k^2_t) \eeq
where
\beq \label{CHI}
\c (f)\,\,=\,\,2 \,\Psi (1) \,\,-\,\,\Psi ( f ) \,\,-\,\,\Psi ( 1 - f )\,\,
\eeq
and
$$\Psi ( f )\,\,=\,\,\frac{ d\,\, \ln \Gamma ( f )}{ d\,\, f}\,\,,$$
$\Gamma ( f ) $ is the Euler gamma function.

\subsubsection{ The general solution of the BFKL equation.}
 From \eq{EIGENVALUE} we can easily  find the general solution of the
BFKL equation using double Mellin transform:
\beq \label{MELLIN}
\p (y, k^2_t)\,\,= \,\,\int \,\,\frac{d \omega}{ 2 \pi \,i}
e^{\omega\,y} \p ( \omega, k^2_t)\,\,=\,\,\int \frac{d
\omega d f}{( 2 \pi \,i)^2} e^{\omega\,y} \p_f ( k^2_t )\, C ( \omega, f )
\eeq
where the contours of integration over $\omega$ and $f$ are situated to the
right of all singularities of $\p ( f ) $ and $ C ( \omega, f ) $.
For $C( \omega, f )$ the equation reads
\beq \label{OMEGALI}
\omega\,C ( \omega, f )\,\,=\,\,\frac{\as N }{ \pi}\,\c ( f ) \, C ( \omega,
f )\,\,.
\eeq
Finally, the general solution is
\beq \label{GENSOL}
\p (y, k^2_t ) \,\,=\,\,\frac{1}{  2 \pi \,i }\,\int \,d f\,e^{\,\frac{\as
N}{\pi}\,
\c ( f ) y \,+\, ( f - 1 )\,r}  \tilde \p ( f )
\eeq
where $\tilde \p ( f ) $ should be calculated from the initial condition at
$ y\,=\,y_0 $ and $ r\,=\, ln \frac{k^2_t}{q^2_0}$ ( $q^2_0$ is the value
of virtuality from which we are able to apply perturbative QCD ) .

\subsubsection{Anomalous dimension from the BFKL equation.}
 We can solve \eq{OMEGALI} in a different way and find $ f \,=\,\g (\omega)$.
  $\g (\omega)$ is the anomalous dimension in LL(log (1/x)A
\footnote{From \eq{MELLIN} one can notice that  moment variable N defined
such that N = $\omega$ + 1.}
and for $\g( \omega ) $ we have the following series \cite{LIANDI}
\beq \label{LIANDI}
\g ( \omega ) \,\,=\,\,\frac{\as N }{\pi}\cdot \frac{1}{\omega} \,\,+\,\,
\frac{2  \as^4 N^4 \zeta ( 3 )}{\pi^4} \cdot\frac{1}{\omega^4}\,\,+\,\,O
( \frac{\as^5}{ \omega^5} )
\eeq
The first term in \eq{LIANDI} is the anomalous dimension of the GLAP equation
\cite{GLAP} in leading order of $\as$ at $\omega \,\ra\,0$, which gives the
solution for the structure function at $ x\,\ra\,0 $ and corresponds to so
called double log approximation of perturbative QCD ( DLA).The  DLA sums
 the contributions of the order $ ( \as \,log (1/x)\,log (Q^2/q^2_0) )^n$
 in the perturbative series of \eq{PERSER}.

However, we would like to stress that \eq{LIANDI} is valid only at fixed
 $\as$ while the anomalous dimension in the GLAP equation can be calculated
for running $\as$. It means that we have to introduce the running $\as$ in
the BFKL equation to achieve a matching with the GLAP equation in the region
where $\omega \,\ll\,1$ and $ \frac{\as}{\omega} \,< \, 1$.

The second remark is the fact that we can trust the series of \eq{LIANDI}
only for the value of $\oa \,\gg \,\ol$ , where
\beq \label{OMEGALIP}
\ol\,\,=\,\,\frac{\as \,N}{\pi}\cdot \c (\,\frac{1}{2}\,)\,\,=\,\,\frac{ 4 N
\,ln 2\,\as}{\pi}\,\,.
\eeq
In vicinity $ \oa\,\ra \ol $ we have the following expression for $\gamma(
\oa ) $:
\beq \label{GAMMAATOL}
\gamma ( \oa )\,\,=\,\,\frac{1}{2}\,\,+\,\,\sqrt{\frac{
\oa\,\,-\,\,\ol}{\frac{\as N}{\pi} \,14 \zeta ( 3 )}}\,\,.
\eeq
Substituting \eq{GAMMAATOL} in \eq{MELLIN} we have
\beq \label{SOLATOL}
\p (y, k^2_t)\,\,= \,\,\int \,\,\frac{d \omega}{ 2 \pi \,i}\,\,
e^{\omega\,y} \p ( \omega, k^2_t)\,\,=\,\,\int
\,\,\frac{d \omega}{ 2 \pi \,i}\,\,
e^{\omega\,y \,\,+\,\,(\,\gamma( \oa )\,-\,1\,)\,r} \ti \p ( \omega )
\,\,=
\eeq
$$
\int \,\,\frac{d \omega}{ 2 \pi \,i}\,\,
e^{(\,\omega\,-\,\ol\,)\,y \,+\,(\,-\frac{1}{2} \,+\,\sqrt{\frac{
\oa\,\,-\,\,\ol}{\frac{\as N}{\pi} \,14 \zeta ( 3 )}}\,\,\,)\,\,r} \,\ti  \p (
\omega)\,\,.
$$
Evaluating the above integral using saddle point approximation we obtain
\beq \label{SADDLEOM}
\oa_{S} \,\,=\,\,\ol \,\,+\,\,\frac{1}{\frac{\as N}{\pi} \,14 \zeta ( 3
)}\cdot\frac{r^2}{4 \,y^2}\,\,
\eeq
which gives the answer:
\beq \label{SOLDIFF}
\p (y, k^2_t )\,\,=\,\,\frac{1}{\sqrt{k^2_t\,q^2_0}}\cdot\ti \p (\oa_{S})\cdot
\sqrt{\frac{2\,\pi\,(\,\oa_{S} -\ol\,)}{y}}\cdot e^{ \ol\,y\,-\,\frac{ln^2
\frac{k^2_t}{q^2_0}}{\frac{\as N}{\pi} \,28 \zeta ( 3
)\,y}}\,\,.
\eeq
We can trust this solution in the kinematic region where $ ( \,ln
\frac{k^2_t}{q^2_0} \,)^2\,\leq\,\frac{\as N}{\pi} \,28 \zeta ( 3
)\,y$. The solution of \eq{SOLDIFF} illustrates one very important property
of the BFKL equation, namely $k^2_t$ can be not only large, but with the
same probability it can  also be very small.
It means that if we started with sufficiently big value of virtuality $q^2_0$
at large value of $y\,=\,ln ( 1/x )$ due to evolution in $y$ the value of
$k^2_t$  could be small ( $k_t \,\sim \L$, where $\L$ is QCD scale ).
Therefore,
the BFKL equation is basically not perturbative  and the worse thing, is
that
we have not yet  learned what kind of assumption about the confinement
has been made in the BFKL equation.

Our strategy for the further presentation is to keep $k^2_t \,>\,q^2_0$ and to
study what  kind of nonperturbative effect we can expect on  including the
running $\as$ in the BFKL equation, as well as  changing  the value of
$\ol$ in the series of \eq{LIANDI}.

 \subsubsection{The bootstrap property of the BFKL equation.}
We have discussed the BFKL equation for the total cross section,
however this equation can also  be proved for the amplitude at transfer
momentum
$q^2 \,\neq\,0$, and not only for colourless state of two gluons in t-channel.
The general form of the BFKL equation  in
$\omega$ - representation  looks as follows \cite{BFKL} ( see \eq{MELLIN}:
 \beq \label{BFKLGEN}
(\, \omega\,\,-\,\,\omega^G ( k^2_t )\,\,-\,\,\omega^G ( ( q - k_t)^2 )\, )
\phi ( \omega, q, k_t )\,\,=\,\,\frac{ \as}{2  \pi} \,\l_R \int \frac{d^2
k'_t}{ \pi} K (q, k_t, k'_t ) \phi (\omega, q, k'_t)\,\,,
\eeq
where the kernel $K(q, k_t, k'_t )$ describes only gluon emission and
\beq \label{KERNELQ}
K ( q, k_t, k'_t )\,\,=\,\,\frac{k^2_t}{ ( k_t - k'_t )^2 \,k'^2_t }\,\,+\,\,
\frac{ ( q - k_t )^2}{ ( k_t - k'_t )^2 \,( q - k'_t )^2 }\,\,-\,\
\frac{q^2_t}{ ( q_t - k'_t )^2 \,k'^2_t }\,\,,
\eeq
$\l_R$ is colour factor wher  $ \l_1 = 2 \l_8 = N $ for singlet and
$ ( N^2 - 1 )$ representations of colour SU(N) group and $\omega^G ( k^2_t
)\,\,=\,\, \a^G ( k^2_t )\,-\,1$ ( see \eq{REGGE} ).

The bootstrap equation means that the solution of the BFKL equation for
octet colour state of two gluons ( for colour SU(3) ), should give the
reggeized gluon with the trajectory $\a^G (k^2_t)$ ( or $\omega^G (k^2_t )$
) given by \eq{REGGE}. The fact that the  gluon becomes a  Regge pole have
been
shown by us in the example of the next to the Born Approximation, and has
been used to get the BFKL equation in the singlet state. It means that the
solution of the BFKL equation in the octet state should have  the  form of a
Regge pole :
\beq \label{RP}
\p (\omega,q,k_t)\,\,=\,\,\frac{Const}{ \omega\,\,-\,\,\omega^G ( q^2 )}
\eeq
Assuming \eq{RP}, one arrives to the following  bootstrap equation:
\beq \label{BOOTSTRAP}
\, \omega^G ( q^2 )\,\,-\,\,\omega^G ( k^2_t )\,\,-\,\,\omega^G ( ( q -
k_t)^2 )\, \,\,=\,\,\frac{ \as}{2  \pi} \,\l_8 \int \frac{d^2
k'_t}{ \pi} K (q, k_t, k'_t ) \,\,,
\eeq
It is easy to check that the trajectory of \eq{REGGE} satisfies this equation.
It is interesting to mention that we can use \eq{BOOTSTRAP} to reconstruct
the form of the kernel $K (q, k_t, k'_t )$,  if we know the expression for
the trajectory ( see ref.\cite{BRAUN} ).
\subsection{The BFKL equation with running $\as$ in the lowest order.}
 To take into account the running $\as$ we shall deal with QCD having large
number of massless fermions $N_f$. In this case we can only insert the chain of
fermions bubbles in a  gluon line in the Feynman diagrams to calculate the
contributions of running $\as$ \cite{BLM}. Indeed, each such bubble gives the
contribution of the order of $N_f \as (\m^2)$, where $\m^2$ is the
renormalization scale and there are no other contributions  of the same order.
Due to the renormalization property of the QCD we have to replace $N_f$ by
$ \,-\,\frac{3}{2}\,\,b \,\,=\,\,-\,[\,\frac{11}{2} N \,-\,N_f\,] $ in the
final
answer, to get the correct contribution of running $\as$ in our problem.

For example, in the Born Approximation for quark - quark total cross section
 we have to replace the diagram of Fig. 2.1 by the sum of diagrams of Fig.
2.8 , inserting fermion bubbles in two gluon lines in t-channel.
 Such a procedure leads to the answer in Born Approximation:
\beq \label{BARUN}
2 \Im{A^{BA}(s,t = 0 )}
\,\,=\,\,s \frac{C^2_2}{ N^2 - 1} \,\int\,\frac{ \as^2 ( k^2_t ) d^2 k_t}{
k^4_t}\,\,.
\eeq
In the next order  to Born Approximation  we have to:

1.   Introduce the fermion
loops in the amplitudes for production  of one additional gluon (see Fig.
2.2 ),
and to the virtual correction diagrams ( see Fig. 2.3) in \eq{NEBA}. Which
means
that one should calculate the sets of diagrams of Fig. 2.9 and Fig. 2.10
respectively.

2. Take into account  the
additional contribution of produced quark - antiquark pair in the final state
( see Fig.2.11) in the unitarity equation ( see \eq{NEBA} ).

 Using the technique developed in ref. \cite{LIFA} we are able to calculate
the diagrams of Figs. 2.9 and 2.11, however we have not yet finished these
calculations and we intend  publishing  them elsewhere.  Instead of the
direct calculations of these sets of diagrams we
chose the alternative approach, namely to calculate the  contribution  due to
running $\as$ only to the virtual correction diagrams of Fig. 2.10,  and
 to use the bootstrap equation (~\ref{BOOTSTRAP})  to reconstruct the
form of the kernal of the BFKL equation with running $\as$.

Repeating all technoques that we have used in \eqs{DISRE}{REGGE} we end up
with
the function $ \Sigma (k^2_t)$ which is equal the diagram of Fig. 2.8 at
$t\,=\, - k^2_t$
\footnote{We absorb one power of $\as$ in the definition of $\Sigma ( k^2_t)$
with respect to \eq{RE} to make easier the counting of  the power of $\as$.} :
 \beq \label{SIGMARUN}
\Sigma ( k^2_t ) \,\,=\,\, \frac{4  }{\pi^2}\,\int \frac{ \as ( k'^2_t )
\,\as( (\,k_t \,-\,k'_t\,)^2) \, d^2 k'_t}{ ( \,k_t - k'_t \,)^2\,k'^2_t}\,\,.
\eeq
Comparing \eq{SIGMARUN} with the Born Approximation ( \eq{BARUN} and Fig. 2.8 )
we get
\beq \label{REGGERUN}
\a^G ( k^2_t )\,-\,1\,=\,\omega^G ( k^2_t )\,\,=\,\,-\,{\l_8}{ 2\,\pi}\,\int
\frac{ d^2
k'_t}{\pi} \cdot \frac{\as( k'^2_t)\,\as( (\,k_t\,-\,k'_t\,)^2 )}{ \as( k^2_t)}
\cdot \frac{k^2_t}{ k'^2_t \,(\,k_t\,-\,k'_t\,)^2}\,\,.
\eeq
 Therefore, we have  established the form of the trajectory
of reggeized gluon in the next to Born Approximation calculation. In the next
subsection we will show if we assume that the
reggeization of the gluon is so general property of the BFKL equation that
it
 also  holds in the case of running $\as$, than the knowledge of $\omega(
k^2_t)$ is enough to
get the kernel for the BFKL equation with running $\as$.
\subsection{The BFKL equation with running $\as$ from the bootstrap equation.}
The bootstrap equation ( see \eq{BOOTSTRAP} )  gives the relation between gluon
trajectory and the kernel of the BFKL equation. It is complicated functional
- integral equation which relates $\omega^G (k^2_t)$ and $K ( q,k_t,k'_t )$,
but
in ref. \cite{BRAUN} it was found  that the solution of this equation can be
parametrized  in the form:
\beq \label{BFKLKER}
K ( q, k_t, k'_t )\,\,=\,\,
\frac{\eta ( k_t )}{ \eta ( k'_t )\,\eta( k_t - k'_t )}\,\,+\,\,
\frac{\eta ( q_t - k_t )}{ \eta ( q_t - k'_t )\,\eta( k_t - k'_t )}\,\,-\,\,
\frac{\eta ( q )}{ \eta ( k'_t )\,\eta( q_t -  k'_t )}\,\,,
\eeq
where the function $\eta$ is related to the trajectory $\omega^G$ by a
nonlinear integral equation:
\beq \label{RETA}
\omega^G ( q )\,\,=\,\,-\,\frac{\l_8}{2\,\pi}\,\int \frac{d^2 k'_t}{ \pi} \,
\frac{\eta ( q )}{ \eta ( k_t )\,\eta ( k_t - k'_t )}\,\,.
\eeq
One can check that \eqs{BFKLKER}{RETA} satisfy the bootstrap equation
(~\ref{BOOTSTRAP} ).

Comparing the equation for the gluon trajectory with running $\as$ that we have
calculated in the previous subsection ( see \eq{REGGERUN} ) we get
\beq \label{ETA}
\eta ( q) \,\,=\,\,\frac{ q^2_t}{\as ( q^2_t )}\,\,.
\eeq
Finally, the BFKL equation with running $\as$ for the total cross section (
$q^2_t\,=\,0$ ) has the form:
\beq \label{BFKLRUN}
 \frac{d \p (y,  k^2_t )}{d y}\,\,=\,\,\frac{ N}{\pi} \,\int d k'^2_t\,\,
 K (k_t,
k'_t ) \p ( y, k'^2_t )\,\,,
\eeq
where   we can get the expression for kernel $K$  substituting \eq{ETA} into
\eq{BFKLKER}\footnote{We absorbed the ratio $\frac{k^2_t}{k'^2_t}$ in the
definition of $\p$ to make a  clear correspondence with the BFKL equation at
fixed $\as$ (see \eq{BFKL} ). } :
 \beq \label{FINKER} K ( k_t, k'_t ) \p ( k^2_t )
\,\,=
\eeq
$$
\,\, \frac{\as( k'^2_t)\,\as (\, ( k_t - k'_t )^2\,)}{\as( k^2_t )}\,\{
\frac{1}{ (\,k_t\,-\,k'_t\,)^2}\cdot \p ( k'^2_t ) \,\,-
\,\,
\frac{k^2_t}{(\,k_t\,-\,k'_t\,)^2\,[\, (\,
k_t\,-\,k'_t\,)^2\,+\,k'^2_t\,]}\cdot \p ( k^2_t )\,\}
$$
The initial condition for \eq{BFKLRUN} in the case of quark - quark
interaction looks as follows:
\beq \label{INCON}
\p^{(1)}\,\,=\,\,\frac{C_2 \as( k^2_t )}{k^2_t}
\eeq
\subsection{ The matching with the GLAP  evolution equation at $x \,\ra\,0 $.}
In this subsection we are going to demonstrate that \eq{BFKLRUN} naturally
gives the double log limit of the GLAP equation in the kinematic region where
not only $\as\,\, log (1/x) \,\sim \,1$, but also the virtuality $k^2_t$ is
large
enough ( $\as\,\, log ( k^2_t/q^2_0) \,\sim\,1$ ). This limit corresponds the
integration over $k'^2_t\,\,\ll\,\,k^2_t $ in \eq{BFKLRUN}. The BFKL equation
(~\ref{BFKLRUN}) has a very simple form in this kinematic region:
\beq \label{DLABFKL}
\frac{ d \p (y, k^2_t )}{ d y }\,\,=\,\,\frac{N}{\pi} \frac{1}{
k^2_t}\cdot\int^{k^2_t} \as (k'^2_t )\, \p( y, k'^2_t)\, d k'^2_t\,\,.
\eeq
It should be stressed that the gluon structure function can be written
uising the  function $\p$ in the following way \cite{TRAFACT}:
\beq \label{GLOUNSTR}
\as( Q^2) x G (x, Q^2 )\,\,=\,\,\int^{Q^2} \as( k^2_t ) \cdot \p (ln (1/x),
k^2_t )\, d k^2_t\,\,.
 \eeq
Using this equation we can rewrite \eq{DLABFKL}  in the form of differential
equation with respect to $x G (x, Q^2 ) $:
\beq \label{DIFDLA}
\frac{ \pa ( x G (x, Q^2 )  )}{ \pa ln (1/x)}\,\,=\,\,\frac{N}{\pi}
\cdot \int^{Q^2} \frac{\as ( k^2_t ) d k^2_t}{k^2_t} \cdot ( x G (x, k^2_t
))\,\,.
\eeq
Taking into account the explicit form of running $\as$, namely
\beq \label{RUNAS}
\as( k^2 ) \,\,=\,\,\frac{ 2\,\pi}{ b\, ln \frac{ k^2}{\L^2}}
\eeq
we can derive the double differential equation for $ x G (x, Q^2 )$ which
looks as follows:
\beq \label{DLADUBDIF}
\frac{\pa^2 ( x G (x, Q^2 ) )}{ \pa ln(1/x) \,\pa \xi}\,\,=\,\,\frac{ 2 N}{b}
 ( x G(x, Q^2 ) )\,\,,
\eeq
where $ \xi\,\,=\,\,ln ln (Q^2/\L^2) $ and $ b\,=\,\frac{3}{2}\,(\, 11 N \,-\,2
N_f\, )$.

This equation is the GLAP equation in the region  of low x. In other words it
is the equation which we can get from the anomalous dimension of \eq{LIANDI}
taking into account only the first term with running $\as$.

\section{The first correction to the BFKL equation due to running $\as$.}
\subsection{ General formula.}
As was discussed the BFKL equation describes a generalized ``ladder"
diagrams ( see Fig. 2.7 ). In this section we are going to calculate the
first correction due to running $\as$ in such a ``ladder". The procedure
how to take into account  this correction is very simple, namely we have to
insert the kernel $ K ( k_t, k'_t )$ of \eq{FINKER} in one cell of the
``ladder",  while in all other cells the kernel remains the kernel of the
BFKL equation with fixed $\as$ ( see \eq{LIKER} ). Fig. 3.1 illustrates
this procedure and gives the graphical picture for the expression that will
follow below.
We will  denote $ K_r $  the
kernel of the BFKL equation with running $\as$ ( see \eq{FINKER}
) and use  the notation $K$ for the kernel of the BFKL equation with fixed
$\as$ so as  to avoid
 any misunderstanding  in further presentation.
 The analytic expression for the diagram of Fig. 3.1  looks as follows:
\beq \label{ANFIRUN}
\p^{[1]} ( y, Q^2 )\,\,=\,\,\int d y'\, \frac{d^2 q_t }{\pi} \,\frac{d^2
q'_t}{\pi}\,\, \p ( y - y', Q^2, q^2_t )\,\, K_r ( q_t, q'_t ) \,\,\p (y' -
y_0 , q'^2_t, q^2_0 )\,\,.
\eeq
Substituting in \eq{ANFIRUN} the solution of the BFKL equation with fixed
$\as$ in the form:
\beq \label{SOLBFKLFIXGAMMA}
\p ( y, Q^2, q^2_t )\,\,=\,\,\int \,\frac{d \omega }{ 2
\pi\,i}\,\,\,e^{\omega\,y \,+\, (\,\gamma ( \omega )\,-\,1\,)\,ln
\frac{Q^2}{q^2_t} \,}\,\,\tilde \p ( \omega )\,\,,
\eeq
where $\gamma ( \omega ) $ is given by \eq{LIANDI}, one can get the following
answer for $\p^{[1]}$:
\beq \label{PHIONE}
\p^{[1]} ( y, Q^2 )\,\,=\,\,\frac{N}{\pi\,\,Q^2 }\,\int \frac{d \omega_1 d
\omega_2}{ ( 2 \pi \, i )^2}\,\frac{ d^2 q_t}{ \pi} \,\frac{d^2 q'_t}{ \pi}\,\,
\tilde \p_1 ( \omega_1 )\, \tilde \p_2 ( \omega_2 ) \,e^{\omega_1 ( y - y'
)\,+\,\omega_2 y'}
\eeq
$$\{\,
e^{\gamma ( \omega_1 )\,\,ln \frac{Q^2}{q^2_t}}
\,e^{\gamma ( \omega_2 )\,\, ln \frac{q'^2_t}{q^2_0}}
\,\cdot \frac{\as ( q^2_t ) \,\as (\, ( q_t\,-\,q'_t )^2\,) }{ \as ( q'^2_t )}
\cdot \frac{1}{ q'^2_t\,(\,q_t\,-\,q'_t\,)^2}\,\,-
$$
$$\,\,
e^{ \gamma ( \omega_1 )\,ln \frac{Q^2}{q^2_t}}
\,e^{ \gamma ( \omega_2 )\,ln \frac{q^2_t}{q^2_0}}
\,\cdot \frac{\as ( q'^2_t ) \,\as (\, ( q_t\,-\,q'_t )^2\,) }{ \as ( q^2_t )}
\cdot \frac{1   }{ q'^2_t\,[\,(\,q_t\,-\,q'_t\,)^2
\,+\,q'^2_t\,]\,}\,\,\}\,\,. $$
It should be stressed that $\ti \p_1( \omega_1 )\,\,=\,\,\frac{d
\gamma(\omega_1 )}{ d \omega_1}$ satisfies the initial condition for $\phi (y
- y', Q^2, q^2_t )\,\,=\,\,\delta ( ln ( Q^2 /q^2_t ) $ at $y = y'$ for upper
part of the diagram of Fig. 3.1, while $\ti \p_2 ( \omega_2 )$ can be found
from the initial condition related to distributions of gluons in the hadron
(or in the quark ) at $y' = y_0 $.

After integration over $y'$ one gets $\omega_1\,=\,\omega_2$ and \eq{PHIONE}
can be reduced to the form:
\beq \label{PHIONE!}
\p^{[1]} ( y, Q^2 )\,\,=\,\,
\eeq
$$
\frac{N}{\pi Q^2}\,\int
\frac{d \omega_1 }{ 2 \pi \, i }\, \,\frac{d^2 q'_t}{ \pi q'^2_t}
\tilde \p_1 ( \omega_1 )\, \tilde \p_2 ( \omega_1 ) \,e^{\omega_1\, y \,+\,
\gamma ( \omega_1 )\,ln \frac{Q^2}{q^2_0}}
\,
\,\cdot \frac{1}{\as( q'^2_t )}\,\cdot\,
 \ti K_r (\, \gamma( \omega_1 ); \as( q'^2_t )\, )\,\,,
$$
where the kernel  $ \ti K_r (\, \gamma( \omega_1 ); \as( q'^2_t )\, )$ is
equal:
\beq \label{KERALGA}
 \ti K_r (\, \gamma( \omega_1 ); \as( q'^2_t )\, )\,\,=\,\,\int \,\frac{ d^2
q_t}{\pi}\cdot\frac{\as ( q^2_t ) \,\as (\, ( q_t\,-\,q'_t )^2\,) }
{ (\,q_t\,-\,q'_t\,)^2}\cdot\{\,\, e^{(\,\gamma ( \omega_1 )\,-\,1\,)ln
\frac{q'^2_t}{q^2_t}} \,\,-\,2\,e^{\,ln\frac{q'^2_t}{q^2_t}}\,\,\}
\eeq
\subsection{Simplification of ${\bf \ti K_r }{\bf(\, \gamma( \omega_1 )};
{ \bf \as( q'^2_t )\, )}$.}
We now  discuss the contribution of infrared and ultraviolet
renormalons to the kernel\\ $ \ti K_r (\, \gamma( \omega_1 ); \as( q'^2_t
)\,)$,
but let us first simplify the expression for this kernel.

1. Choosing the renormalization point $\m^2\,=\,q'^2_t$ we can rewrite the
running $\as$ in \eq{KERALGA} in the form:
\beq \label{RAS}
\as ( q^2_t )\,\,=\,\,\frac{\as( q'^2_t )}{ 1\,\,+\,\,\frac{b}{2 \pi }\,\as (
q'^2_t ) \,ln \frac{q^2_t}{q'^2_t} }\,\,=\,\,\frac{\as(
q'^2_t )}{ 1\,\,+\,\,\ti \as (q'^2_t ) \,ln \frac{q^2_t}{q'^2_t} }
\eeq
$$
\as (\,( q_t\,-\,q'_t )^2\,)\,\,=\,\,\frac{\as( q'^2_t )}{ 1\,\,+\,\,\frac{b}{2
\pi } \,\as ( q'^2_t ) \,ln \frac{(\,q_t\,-\,q'_t\,)^2}{q'^2_t} }\,\,=\,\,
\frac{\as( q'^2_t )}{ 1\,\,+\,\,\ti \as ( q'^2_t ) \,
ln \frac{(\,q_t\,-\,q'_t\,)^2}{q'^2_t} }\,\,=\,\,.
 $$
Using \eq{RAS} we can integrate over angle in \eq{KERALGA}:
\beq \label{INTANGLE}
\int d \p \frac{\as (\, ( q_t\,-\,q'_t )^2\,)}{(\,q_t\,-\,q'_t\,)^2}\,\,=\,\,
 \as (q'^2_t) \Sigma^{\infty}_{i= 0}  ( - 1 )^i {\ti \as ( q'^2_t )}^i \frac{
d^i}{( d \m )^i}|_{\m=0}\,\,\int \frac{d \p ( q'^2 )^{- \m}}{
(\,q_t\,-\,q'_t\,)^{ 1 - \m}}\,\,=
\eeq
$$
\,\,\pi\, \as (q'^2_t) \Sigma^{\infty}_{i= 0}  ( - 1 )^i {\ti \as ( q'^2_t
)}^i \frac{
d^i}{( d \m )^i}|_{\m=0}\,\,\frac{1}{|\,q^2_t\,-\,q'^2_t\,|}\cdot [\, \frac{|
\,q^2_t\,-\,q'^2_t\,|}{q'^2_t}\,]^{2\m}\,F( \m, \m, 1, z )\,\,\ra
$$
$$
\ra\,\,\pi
\frac{\as( q'^2_t )}{ 1\,\,+\,\,2\,\ti \as ( q'^2_t ) \,
ln \frac{|\,q^2_t\,-\,q'^2_t\,|}{q'^2_t}
}\,\cdot\,\frac{1}{|\,q^2_t\,-\,q'^2_t\,|}\,\,+\,\,O (z\as^3(q'^2_t) )\,\,,
$$
where $z\,=\frac{q'^2_t}{q^2_t} $ for $q'^2_t\,<\,q^2_t$ and $
z\,=\frac{q^2_t}{q'^2_t} $ for $q^2_t\,<\,q'^2_t$, $F(\m,\m, 1, z)$ denotes the
Gauss hypergeometric function.

The next trick we suggest to work with the product of $\as$ in \eq{KERALGA}:
\beq \label{TRICKALPHA}
\frac{1}{ 1\,\,+\,\,\ti \as (
q'^2_t ) \,ln \frac{q^2_t}{q'^2_t}}\,\cdot\,\frac{1}{ 1\,\,+\,\,2\,\ti \as
( q'^2_t ) \,ln \frac{|\,q^2_t\,-\,q'^2_t\,|}{q'^2_t}}\,\,=
\eeq
$$
\frac{1}{ 1\,\,+\,\,\ti \as (q'^2_t ) \,ln \frac{q^2_t}{q'^2_t}}
\,\,+\,\,
\frac{1}{ 1\,\,+\,\,2\,\ti \as( q'^2_t ) \,ln
\frac{|\,q^2_t\,-\,q'^2_t\,|}{q'^2_t}}
\,\,-\,\,1\,\,+
$$
$$
\,\,2\,\,\frac{{\ti\as}^2( q'^2_t)\,\,ln \frac{q^2_t}{q'^2_t}\,\,ln
\frac{|\,q^2_t\,-\,q'^2_t\,|}{q'^2_t \,}}{ (\,1\,\,+\,\,\ti \as (
q'^2_t ) \,ln \frac{q^2_t}{q'^2_t}\,)\cdot(\, 1\,\,+\,\,2\,\ti \as
( q'^2_t ) \,ln \frac{|\,q^2_t\,-\,q'^2_t\,|}{q'^2_t}\,)}\,\,.
$$
We can neglect the contribution of the last term in \eq{TRICKALPHA}, since it
is
proportional to $\as^2$ and both logs cannot be large simultaneously. We have
checked
numerically that this contribution is really very small, but it is even more
important
for us, that this term do not bring any nonperturbative phenomena in the
question of
the interest, which have not been included in the first three terms.

Using variable $z\,\,=\,\,\frac{q^2_t}{q'^2}$ which we have introduced we can
rewrite the
expression
for kernel $\ti K_r (\, \gamma( \omega_1 ); \as( q'^2_t )\, )$  in the form:
\beq  \label{FKOAL}
\ti K_r (\, \gamma( \omega_1 ); \as( q'^2_t )\, )\,\,=
\,\,\as^2(q'^2_t)\int \,\frac{ d z}
{z\,\,|\,1\,-\,z\,|}\cdot
\eeq
$$
\{\frac{1}{1\,\,+\,\,\ti \as( q'^2_t )\,ln
z}\,\,+\,\,\frac{1}{1\,\,+\,2\,\ti \as (q'^2_t )\,ln
|\,1\,-\,z\,|}\,\,-\,\,1\,\,\}  \,\cdot\,
\{\, e^{(\, 1 -\,\gamma ( \omega_1 )\,)\,\,ln z}
\,\,-\,\,e^{\,ln z}\,\}
$$
\subsection{Infrared Renormalons.}
\subsubsection{$ {\bf \ti K_r (\, \gamma( \omega_1 ); \as( q'^2_t )\, )}$ at
{\bf $q^2_t \,<\, q'^2_t$}\,\,.}
The IR renormalons contribution comes from the kinematic region $q^2_t
\,<\, q'^2_t$. Let us simplify \eq{FKOAL} in this kinematic region introducing
the new variable  $u$\,\, $z\,\,=\,\,e^{u}$:
\beq \label{FKOALSQ}
\ti K_r (\, \gamma( \omega_1 ); \as( q'^2_t )\,
)\,|_{q^2_t\,<\,q'^2_t}\,=\,\,\as^2 (
q'^2_t )\,\,\int^{\infty}_0 \,\,d u\,\, \frac{ 1}{ 1\,-\,e^{-\,u}}\,\cdot
 \eeq
$$
\{ \,\frac{ e^{\,-\,(\,1\,-\,\gamma (\oa_1)\,)\,u}\,\,-\,\,e^{\,-\,u}}{
1\,\,-\,\,\ti \as ( q'^2_t )\,\,u}\,\,+\,\,\frac{2 \ti \as ( q'^2_t )\,\,ln (
\,1\,-\,e^{\,-\,u}\,)\,\cdot \,[\, e^{\,-\,(\,1\,-\,\gamma
(\oa_1)\,)\,u}\,\,-\,\,e^{\,-\,u}\,]}{ 1\,\,-\,\,2\,\as ( q'^2_t) \,\,ln
(\,1\,-\,e^{\,-\,u}\,)}\,\,\}\,\,. $$
Changing the variable of integration in the second term of \eq{FKOALSQ} $
1\,-\,e^{-\,u} \,\,\ra\,\,e^{-\,\frac{u}{2}} $ we can rewrite \eq{FKOALSQ} in
 the form:
\beq \label{FINFKOALSQ}
\ti K_r (\, \gamma( \omega_1 ); \as^2( q'^2_t )\, )|_{q^2_t
\,<\,q'^2_t}\,\,=\,\,\as (
q'^2_t )\,\,\int^{\infty}_0 \,\,d u \,\,\frac{ 1}{ 1 \,-\,\as( q'^2_t )
\,u}\,\cdot
 \eeq
$$
\{\,\,\frac { e^{\,-\,(\,1\,-\,\gamma( \oa_1 )\,)\,u}\,\,-\,\,e^{\,-\,u}
}{1\,\,-\,\,e^{\,-\,u}}\,\,+\,\,\frac{1}{2}\,\frac{ \as ( q'^2_t ) \,u\,\,[
\,(\,1\,\,-\,\,e^{\,-\,\frac{u}{2}}\,)^{ 1\,-\,\gamma ( \oa_1 )}\,\,-\,\,
(\,1\,\,-\,\,e^{\,-\,\frac{u}{2}}\,)\,]}{
1\,\,-\,\,e^{\,-\,\frac{u}{2}}}\,\,\}\,\,.
$$
Eq. (~\ref{FINFKOALSQ}) can be rewritten in more symmetric way for the
difference \beq \label{DIFKOALSQ}
 \Delta \ti K_r \,\,=\,\,\frac{1}{\as( q'^2_t )}\ti K_r (\, \gamma( \omega_1 );
\as( q'^2_t )\, )\,|_{q^2_t \,<\,q'^2_t}\,\,-\,\, \as ( q'^2_t ) \,\chi
( \gamma (\oa_1 ) \,)\,|_{q^2_t \,<\,q'^2_t} \,\,
\eeq
$$
=\,\,\as (
q'^2_t )\,\,\int^{\infty}_0 \,\,d u \,\,\frac{ \as ( q'^2_t ) \,\,u}{ 1 \,-\,
\as(
q'^2_t ) \,u}\,\cdot
\{\,\,\frac { e^{\,-\,(\,1\,-\,\gamma( \oa_1 )\,)\,u}\,\,-\,\,e^{\,-\,u}
}{1\,\,-\,\,e^{\,-\,u}}\,\,
$$
$$
+\,\,\frac{1}{2}\,\frac{ [
\,(\,1\,\,-\,\,e^{\,-\,\frac{u}{2}}\,)^{ 1\,-\,\gamma ( \oa_1
)}\,\,-\,\,(\,1\,\,-\,\,e^{\,-\,\frac{u}{2}}\,)\,]}{
1\,\,-\,\,e^{\,-\,\frac{u}{2}}}\,\,\}\,\,,
 $$
where $\chi( \gamma (\oa_1 )|_{q^2_t \,<\,q'^2_t}$ is the contribution to
the kernel of the BFKL equation ( see \eq{OMEGALI} ) from the kinematic
region $q^2_t\,<\,q'^2_t$.

\subsubsection{IR renormalons: uncertainty of the perturbative series  and
relation to the  shadowing correction.}
 In this subsection we are going to show that $\Delta \ti K_r$ contains terms
of the order of $\as^n\,n!$ which give the natural limit for perturbative
calculation since these contributions  originate from integration over
small value of $q_t$. To see such terms we can expand \eq{DIFKOALSQ} with
respect to $u$ and $e^{ -\,u}$ or $e^{-\frac{u}{2}}$. Indeed, $\Delta \ti K_r $
can be written as the following series:
\beq \label{IRSERIES}
\Delta \ti K_r |_{q^2_t\,<\,q'^2_t} \,\,=\,\,\as ( q'^2_t ) \Sigma_{n = 1}
 \Sigma_{l = 0} \,\,{\ti \as}^n( q'^2_t )\,\,\cdot
\eeq
$$
[\,\int^{\infty}_0
d u \,u^n \,\,\{\,(\,e^{- (\,1\,+\,l\,-\,\gamma( \oa_1
)\,)\,u}\,\,+\,\,\frac{1}{2}\,
\,\frac{\Gamma ( \gamma( \oa_1 ) \,+\,1\,)}{ \Gamma ( \gamma ( \oa_1
)\,+\,l )\,l!}\,e^{\,-\,\frac{( l\,+\,1) \,u}{2}}\,)\,\,-\,( \gamma \,=\,0
)\,\}\,\,]\,\,=
$$
$$
 \frac{2 \pi}{b}\,  \cdot \,\Sigma_{n = 1}\,\Sigma_{l = 0} \,[\,\{\,\,(\,
\frac{
\ti \as (q'^2_t )}{ 1 \,+\,l\,-\,\gamma ( \oa_1 )}\,)^{n + 1}\,\cdot\,\Gamma (
n +
1 ) \,\,+
$$
$$
\,\,\frac{1}{2}\,
\,  \frac{\Gamma ( \gamma( \oa_1 ) \,+\,1\,)}{ \Gamma ( \gamma ( \oa_1
)\,+\,l )\,l!}\,(\,\frac{2\,\ti \as ( q'^2_t )}{l\,+\,1}\,)^{n + 1}
\,\cdot\,\Gamma ( n + 1 )\,\,\}\,\,+\,\,\{ \gamma = 0 \} \,\,]\,\,=
$$
$$
 \,\,\as \,\,\Sigma_{n = 1}\,\,\,\Sigma_{l = 0}\,{\ti \as}^n
\,\,R_{n,l}
 $$
In \eq{IRSERIES} the integral representation for Euler gamma function has been
used, namely
$$ \Gamma ( n + 1 ) \,\,=\,\,\int^{\infty}_0 t^n\,e^{-\,t} d t$$
 Eq. (~\ref{IRSERIES} )  is a typical asymptotic series which
is no longer reliable, starting from critical order $n = N $ which can be
found from the condition:
\beq \label{NCON}
 \as^N ( q'^2_t )R_{N, l }\,\,=\,\,\as^{ N + 1} R_{N + 1, l}\,\,,
\eeq
as has been discussed in the introduction. The answer for the $\Delta \ti K_r$
can be written in the form:
\beq \label{ANDELTA}
\Delta \ti K_r \,\,=\,\,\as \,\Sigma^{N}_{n = 1}\,\,\Sigma_{ l = 0} \as^n
R_{n. l} \,\,\,\pm\,\,\,\as^{n + 2} R_{N + 1, l}\,\,.
\eeq
 One can check that the largest uncertainties originates from $ l = 0$ in the
series of \eq{IRSERIES} which leads to the value of $N$ equal:
$$
N\,\,=\,\,\frac{ 1 \,-\,\gamma( \oa_1 )}{\ti \as ( q'^2_t )}\,\,=\,\,(\,
1\,\,-\,\,\gamma ( \oa_1 )\,)\cdot ln \frac{q'^2_t}{\L^2}  \,\,.
$$
The corresponding uncertainty in the perturbative expansion is given by
\beq \label{UNCERT}
\delta (\, \Delta \ti K_r\,)\,|_{q^2_t \,<\,q'^2_t} \,\,=\,\,\as ( q'^2_t )\,
\sqrt{\pi ( 1 - \gamma ( \oa_1 ) ) \ti \as ( q'^2_t )}\,\,\{\, \frac{\L^2}{
q'^2_t}\,\}^{-\,1 \,+\,\gamma( \oa_1 )}
\eeq
Substituting \eq{UNCERT} in \eq{PHIONE!} one can calculate the uncertainty in
$\p^{[1]}$:
\beq \label{UNCERTPHIONE}
\delta \phi^{[1]}\,\,=\,\,
\frac{N}{\pi Q^2}\,\int
\frac{d \omega_1 }{ 2 \pi \, i }\, \,\frac{d^2 q'_t}{ \pi q'^2_t}
\tilde \p_1 ( \omega_1 )\, \tilde \p_2 ( \omega_1 ) \,e^{\omega_1\, y \,+\,
\gamma ( \omega_1 )\,ln \frac{Q^2}{q^2_0}}\,\,\cdot
\eeq
$$
\as ( q'^2_t )\,
\sqrt{\pi ( 1 - \gamma ( \oa_1 ) ) \ti \as ( q'^2_t )}\,\,\{\, \frac{\L^2}{
q'^2_t}\,\}^{-\,1 \,+\,\gamma( \oa_1 )}\,\,.
$$

The fact that $\delta \p^{[1]}$ is proportional to the QCD scale $\L$
indicates the nonperturbative origin of the uncertainties. The physical
meaning of this phenomena is well known \cite{ZAKHMU}. Indeed, the typical
value of the momentum  ( $ ( q_{eff} )_n $ ) essential in the integral for
$\Gamma ( n )$ in \eq{DIFKOALSQ}, is equal to:
\beq \label{TYPMO}
 ( q^2_{eff} )_n \,\,\approx\,\,q'^2 \,\cdot\,e^{ -\,\frac{n}{1 - \gamma(
\oa_1 ) }} \eeq
and at large value of $n$ this momentum can be very small, so small that we
cannot use the perturbative QCD in our calculations ( $ q_{eff} \,\approx
\,\L$ ).

Of course we cannot  trust the value of \eq{UNCERTPHIONE}, we consider
this equation as the indication that we should examine the nonperturbative
contribution with the same $Q^2$ dependence as  is given by \eq{UNCERTPHIONE}.
In the case of the $e^{+}e^{-}$ - annihilation the uncertainties from IR
renormalons can be absorbed in the nonperturbative value of the gluon
condensate
( see refs. \cite{ZAKHMU} ). What nonperturbative phenomena can be
responsible for the uncertainties in our case is the subject that we now
discuss.

It is very instructive to compare $\delta \p^{[1]}$ with the first diagram (
see
Fig.3.2 ) for the shadowing correction \cite{GLR} \cite{HIGHTW}:
\beq \label{SC}
\p^{[1]}_{SC}\,\,=\,\,\frac{\bar \gamma}{ Q^2 \pi R^2}\cdot \int^y_0\,d
y'\int\,\frac{ d
q'^2}{ q'^4_t}\,\, \p (y - y', Q^2, q'^2_t )\,\, \frac{ \as( q'^2_t )
N}{\pi}\,\,\p^2(y', q'^2_t, q^2_0 )\,\,=\,\,
\frac{\bar \gamma}{Q^2\,\pi\,R^2}\cdot
\eeq
$$
 \int \frac{ d \oa_1 \d \oa'}{ (
2\,\pi )^2 } \,\ti \p_1 ( \oa_1 )\, \ti \p_2 ( \oa_1 - \oa' ) \,\ti \p (\oa' )
\,e^{\oa_1 \,\,y} \,\,\frac{\as ( q'^2_t ) N \, d q'^2}{\pi\,\, q'^4_t} \,\,
e^{ \gamma( \oa_1 ) \,ln \frac{Q^2}{q'^2_t } \,\,+\,\, [\, \gamma ( \oa_1
\,-\,\oa' ) \,\,+\,\,\gamma( \oa')\,]\,ln \frac{q'^2_t}{q^2_0}}\,\,=
$$
$$
\frac{\bar \gamma}{Q^2\,\pi\,R^2}\cdot \int \frac{ d \oa_1 }{
2\,\pi  } \,\ti \p_1 ( \oa_1 )\, {\ti \p}^2_2 ( \frac{\oa_1}{2}  ) \,
\,e^{\oa_1 \,\,y} \,\,\frac{\as ( q'^2_t ) N \, d q'^2}{\pi\,\, q'^2_t} \,\,
\frac{1}{\sqrt{2\pi \gamma''( \frac{\oa_1}{2} )}}\,\,
e^{ \gamma( \oa_1 ) \,ln \frac{Q^2}{q'^2_t } \,\,-\,\, [\,\,2\, \gamma (\frac{
\oa_1}{2} \,\,-\,\,1\,]\,ln \frac{q'^2_t}{q^2_0}}\,\,,
$$
where we substituted $\p$ from  \eq{MELLIN} and integrated over $\oa'$ using
steepest decent method. $\bar \gamma$ is the value of triple "ladder" vertex
and $R^2$ is the two gluon correlation length inside the hadron ( all other
notations are clear from Fig. 3.2 ). One can find more detailed  calculations
of the SC diagram in refs.\cite{GLR}\cite{HIGHTW}.

Comparing \eq{SC} with \eq{UNCERTPHIONE} one can see that the
shadowing correction to the structure function  gives the contribution which
looks very similar to the nonperturbative uncertainty from IR renormalons.
Indeed
$$
\p^{[1]}_{SC} \,\,\propto\,\, \frac{1}{Q^2} \,\,e^{\gamma ( \oa_1 )\,\,ln
\frac{Q^2}{q'^2_t} \,\,+\,\,[\,2\,\gamma( \frac{\oa_1}{2}  )\,-\,1\,]  \,\,ln
\frac{q'^2_t}{q^2_0}}
$$
while
$$
\delta \p^{[1]}\,\,\propto \,\,\frac{1}{Q^2}\,\,e^{\gamma ( \oa_1 )\,\,ln
\frac{Q^2}{q'^2_t} \,\,+\,\,[\,\,\gamma( \oa_1 ) \,-\,1\,]  \,\,ln
\frac{q'^2_t}{q^2_0}}\,\,.
$$
However, one can conclude from the above expression that the SC correction
is always bigger than the uncertainties due to the IR
renormalons contributions,
since $2 \gamma ( \frac{\oa_1}{2} ) \,>\,\gamma( \oa_1 )$.

We therefore conclude. If we take into account such
nonperturbative phenomena as the shadowing correction we can forget about
uncertainties due to the IR renormalons contribution in the kernel of the BFKL
equation with running coupling QCD constant.

However,  we should be very careful with the above statement, because
there is an uncertainty from the second term in \eq{DIFKOALSQ} which gives
the value of $N$
$$
N\,\,=\,\,\frac{1}{2 \ti \as (q'^2_t)}\,\,=\,\,\frac{1}{2}\,\cdot\,ln
\frac{q'^2_t}{\L^2}\,\,.
$$
This $N$ generates the uncertainty in $\p{[1]}$
$$
\delta \p^{[1]}\,\,\propto\,\,\frac{1}{Q^2}\,\,e^{ \gamma( \oa_1
)\,ln\frac{Q^2}{q'^2_t}\,\,+\,\,\frac{1}{2} \,ln\frac{q'^2_t}{q^2_0}}\,\,.
$$
This uncertainty is bigger than the shadowing correction for $2\gamma(
\frac{\oa_1}{2} )\,<\,\frac{1}{2} $. It means that we cannot trust our
calculation of the shadowing correction at $ \omega \,>\, \omega_c $, where
$\omega_c$ can be found from equation $ 2\,\gamma(
\frac{\omega_c}{2} )\,=\,\frac{1}{2}$. In other words, for such value of $\oa$
the nonperturbative corretions to the BFKL equation with running $\as$
can be bigger than the SC contribution. The rough estimate for the value of
$\omega_c$ from the first term of \eq{LIANDI} gives
$\omega_c\,=\,\frac{4\,N\,\as ( q'^2_t )}{\pi}$.

It should be stressed that we have found the contribution to the gluon
structure function which behaves as $\sqrt{\frac{\L^2}{Q^2}}$ and which  does
not appear in the Wilson Operator Product Expansion. It originates from the
small value of the momentum of emitted gluon ( $ q_t - q'_t \,\,\ra\,\,0 $
in Fig. 3.1 ). We suspect that this contribution is closely related to our
fundamental hypothesis on the completeness of the wave functions of the
produced hadrons in the final state of the deeply inelastic processes.
This hypothesis allows us to reduce the calculation of the deep inelastic
structure function to parton ( quark and gluon ) degrees of freedom. However,
the slow hadrons in the current jet can interact with the slow hadrons in
the target jet and such an interaction can violate the assumed
completeness of the wave functions   of produced partons.
We would like to draw your attention to the fact that the ratio of the mass
of the current jet of hadrons to the value of $Q^2$ is equal to
$\frac{\mu}{Q}$, where $\mu$ is the mass of the lightest hadron. However,
we have not found the theoretical description of this nonperturbative
contribution and consider it as a good subject for the further investigation.
 \subsubsection{IR renormalons:
singularities in the Borel plane.}
Let us go back to the asymptotic series of \eq{DIFKOALSQ} and try to
sum this series. To sum asymptotic series, means to find the analytic
function which has the same expansion as the asymptotic series.
The general procedure of how to guess the  analytic function is to use
 the Borel representation for the divergent, perturbation expansion
for
 $$
\Delta \ti K_r\,|_{ q^2_t\,<\,q'^2_t} ( \as (q'^2_t ) )\,\,=
\,\,\as \,\,\Sigma_{n = 1}\,\Sigma_{l = 0}\,{\ti \as}^n
\,R_{n,l}\,\,.
$$
Instead of this expansion we define the function:
\beq \label{BORKSQ}
\Delta K^{B}_r \,|_{ q^2_t\,<\,q'^2_t}\,\,=\,\,\Sigma_{n = 1}\,\Sigma_{l
= 0}\,
\,R_{n + 1,l}\,\,\frac{b^n}{n!}\,\,.
\eeq
It is widely believed that this series has a finite radius of
convergence in the $b$ - plane ( see refs.\cite{ZAKHMU} for detail
discussions how it works in the case of $e^{+}e^{-}$ - annihilation ).
$\Delta K^{B}_r $ is the Borel function corresponding to $ \Delta \ti
K_r\,|_{ q^2_t\,<\,q'^2_t} ( \as (q'^2_t ) )$ and we  get
\beq \label{BORALPHA}
\Delta \ti K_r\,|_{ q^2_t\,<\,q'^2_t} ( \as (q'^2_t ) )\,\,-\,\,\Delta
\ti K_r\,|_{ q^2_t\,<\,q'^2_t} ( 0 )\,\,=\,\,\int^{\infty}_0\,\,db\,\,
\Delta K^{B}_r \,|_{ q^2_t\,<\,q'^2_t}\,\,e^{-\,\frac{b}{\ti \as}}\,\,.
\eeq
Eq. (~\ref{BORALPHA} ) is certainly correct order by order in
perturbation theory, but in general there are two known  reasons why
this
cannot be true for  such asymptotic series.

1.The $b$ - integral in \eq{BORALPHA} does not converge at $b
\,\ra\,\infty$ \cite{BINFTY}. However, this singular behaviour comes from
the mass spectrum in the exact expression for the physical observable and is
irrelevant for the function
 $\Delta \ti K_r\,|_{
q^2_t\,<\,q'^2_t} ( \as (q'^2_t ) )$ for which we have only a  perturbative
expression.

2. There are singularities of $\Delta K^{B}_r$ on the positive real $b$ - axis
making the integral in \eq{BORALPHA} ambiguous \cite{BINFTY} \cite{BNEG} .
Indeed, substituting $R_{n.l}$
from \eq{DIFKOALSQ} in \eq{BORKSQ} one can see that $\Delta K^{B}_r$ has
the simple poles in $b$
$$
\frac{1}{ b \,\,-\,\,b_{0l}}
$$
where
\beq \label{BOLSQ}
b_{0l}\,\,=\,\, a^{(i)}
\eeq
$$
a^{(1)}\,\,=\,\, 1\,\,+\,\,l\,\,-\,\,\gamma( \oa_1 )\,\,;
$$
$$
a^{(2)}\,\,=\,\,1 \,\,+\,\,l\,\,;
$$
$$
a^{(3)}\,\,=\,\, \frac{1\,\,+\,\,l}{2}\,\,;
$$
Therefore the singularities in the Borel plane are  as shown in Fig. 3.3
( we want to mention that we introduced the Borel transform with respect
to $\ti \as$ but not with respect to $\as$ ).

These singularities lead to ambiguities in performing of the Borel integral.
However we have discussed these ambiguities in the previous section and
drove to the conclusion that the uncertainties related to them
are  smaller than the contributions of the shadowing corrections.
It means that we can absorbed all uncertainties related to errors that we
can make performing the Borel integral in the vicinities of the the poles
on positive $b$ - axis in the nonperturbative gluon correlation (
$R^2$ ) length in the expression for the SC correction ( see \eq{SC} ).

We choose the following prescription for the definition of the integral
\beq \label{BORINTSQ}
\int^{\infty}_0 \,\,\frac{d b \,\,e^{-\,\frac{b}{\ti \as}}}{b\,\,-\,\,b_{0l}}
\,\,=\,\,lim\,|_{\eps\,\ra\,0}\,\,\{\int^{b_{0l}\,-\,\eps}_0\,\frac{d b
\,\,e^{-\,\frac{b}{\ti
\as}}}{b\,\,-\,\,b_{0l}}\,\,+\,\,\int^{\infty}_{b_{0l}\,+\,\eps},\frac{d b
\,\,e^{-\,\frac{b}{\ti \as}}}{b\,\,-\,\,b_{0l}}\,\,\}\,\,=
\eeq
$$
\,\,=\,\,-\,\,e^{-\,\,\frac{b_{0l}}{\ti \as}}\,\,Ei \( \,\frac{b_{0l}}{\ti
\as}\,\)\,\,. $$
The $Ei(x)$ is very good analytic function and it solves our task: to find the
analytic function which has the same expansion  as our perturbative series.
\subsubsection{${\bf \Delta \ti K_r}$ for ${\bf q^2_t\,<\,q'^2_t}$.}
Finally the answer for $\Delta \ti K_r$ in the region  $q^2_t\,<\,q'^2_t$
is:
\beq \label{DELTAKSQ}
\Delta \oa_{IR}\,\,=\,\,\Delta \ti K_r \,|_{q^2_t\,<\,q'^2_t}\,\,=\,\,
\frac{\as}{\ti \as}\,\cdot\, \Sigma^{\infty}_{n = 0}\,\cdot
\eeq
$$
\,\{ \,\,\( e^{-\,\frac{1 + n}{\ti \as}}\,\,Ei
\( \frac{1 + n}{\ti \as} \)  \,\,-\,\,\frac{\ti \as}{ n + 1 }\,\,-\,\,
e^{-\, \frac{ 1 + n - \gamma( \oa_1 )}{\ti \as}}\,\,Ei \(  \frac{ 1 + n -
\gamma( \oa_1 )}{\ti \as} \) \,\,+\,\,\frac{\ti \as}{ n  + 1 - \gamma(
\oa_1 )}\,\,\) \,\,+
$$
$$
\,\,\frac{1}{2} \,\,\,\frac{\Gamma (
\gamma ( \oa_1 ) + 1 )}{ \Gamma ( \gamma ( \oa_1 ) + 1 ) \,\,n!}\,\(
\frac{ 2 \ti \as}{ n + 1 } \,\,-\,\,e^{- \frac{n + 1}{2 \,\ti \as}}\,\,Ei
\( \frac{n + 1}{2 \,\ti \as} \)\,\,\)\,\,\}
$$
The series of \eq{DELTAKSQ} is well convergent,  since the general term
of this expansion at any given value of $\as$ falls  as $1/n^2$.
Figs. 3.4 and 3.5 show the numerical estimates for $\Delta \ti K_r$
 versus $\gamma( \oa_1)$  at fixed $\ti \as$ = 0.2 ( Fig. 3.4 ) and versus
$ x = \frac{1}{\ti \as}$ at  given $\gamma\, = \,\frac{1}{2} $. We would
like to draw your attention to the fact that for $\ti \as < 0.33$ the
correction to the value  of the BFKL kernel with fixed $\as$ ( \,$\as \chi (
\frac{1}{2} )$\,) is smaller than 12\% ( $\frac{\Delta K_r}{ \as \chi (
\frac{1}{2} )}\,<\,12\%$).

\subsection{Ultraviolet Renormalons}
\subsubsection{ $ {\bf \ti K_r}{\bf (\, \gamma( \omega_1 )};{\bf \as( q'^2_t
)\, )}$ at ${\bf q^2_t \,>\, q'^2_t}$\,\,.}
 In this subsection we are going to write down the expression for kernel
$K_r$  in the kinematic region $q^2_t \,>\,q'^2_t$ which is related to the
contribution of so called ultraviolet renormalons. Using all techniques  of
section 3.2  and introducing variable $z = e^{-\,u} $ we can rewrite \eq{FKOAL}
for $ u > 0 $ in the form:
\beq \label{FKOALLQ}
\ti K_r (\, \gamma( \omega_1 ); \as( q'^2_t )\,
)\,|_{q^2_t\,>\,q'^2_t}\,=\,\,\as^2 (
q'^2_t )\,\,\int^{\infty}_0 \,\,d u\,\, \frac{ 1}{ 1\,-\,e^{-\,u}}\,\cdot
 \eeq
$$
\{ \,\frac{ e^{\,-\,\gamma (\oa_1)\,\,u}\,\,-\,\,e^{\,-\,u}}{
1\,\,+\,\,\ti \as ( q'^2_t )\,\,u}\,\,-\,\,\frac{2 \ti \as ( q'^2_t )\,\,ln (
\,e^{\,u}\,- \,1\,)\,\cdot \,[\, e^{\,-\,\gamma
(\oa_1)\,u}\,\,-\,\,e^{\,-\,u}\,]}{ 1\,\,+\,\,2\,\ti \as ( q'^2_t )\,\,ln
(\,e^{\,u}\,-\,1\,)}\,\,\}\,\,. $$
  In the second term of \eq{FKOALLQ} we change variable of integration $e^u
-1 \,\ra e^{\frac{u}{2}}$ and the difference $\Delta K_r$ can be rewritten in
the form:
\beq \label{DIFFKOALLQ}
\Delta \ti K_r \,\,=\,\,\frac{1}{\as( q'^2_t )}\ti K_r (\, \gamma( \omega_1 );
\as( q'^2_t )\, )\,|_{q^2_t \,>\,q'^2_t}\,\,-\,\, \as (q'^2_t ) \,\chi
( \gamma (\oa_1 ) \,)\,|_{q^2_t \,<\,q'^2_t} \,\,
\eeq
$$
=\,\,\as (
q'^2_t )\,\,\int^{\infty}_0 \,\,d u \,\,\frac{ \as ( q'^2_t ) \,\,u}{ 1 \,+\,
\as(
q'^2_t ) \,u}\,\cdot
\{\,\,\frac { e^{\,-\,\gamma( \oa_1 )\,u}\,\,-\,\,e^{\,-\,u}
}{1\,\,-\,\,e^{\,-\,u}}\,\,
$$
$$
+\,\,\frac{1}{2}\,\cdot [
\,(\,1\,\,+\,\,e^{\,\,\frac{u}{2}}\,)^{ -\,\gamma ( \oa_1
)}\,\,-\,\,1\,] \,\,\}\,\,.
$$
In the same way as has been done for the region $q^2_t \,<q'^2_t$ we can
expand the above expression with respect to $e^{ - u}$ or $e^{-\, \frac{u}{2}}$
and derive  the following series:
\beq \label{SERDIFFKOALLQ}
\Delta \ti K_r |_{q^2_t\,<\,q'^2_t} \,\,=\,\,\as ( q'^2_t ) \Sigma_{n = 1}
 \Sigma_{l = 0} \,\,(\,{\,-\,\ti \as}\,)^n( q'^2_t )\,\,\cdot
\eeq
$$
[\,\int^{\infty}_0
d u \,u^n \,\,\{\,\,e^{\,-\,( \,l\,+\,\gamma( \oa_1
)\,)\,u}\,\,+ \,\,e^{\,-\,(\,1\,+\,l\,)\,u} \,\,+\,\,\frac{1}{2}\,
\,( - 1 )^l \frac{\Gamma ( \gamma( \oa_1 ) \,+\,1\,)}{ \Gamma ( \gamma ( \oa_1
)\,+\,l )\,l!}\,e^{\,-\,\frac{( l\,+\,1) \,u}{2}}\,\}\,\,]\,\,=
$$
$$
 \frac{2 \pi}{b}\,  \cdot \,\Sigma_{n = 1}\,\Sigma_{l = 0} \,[\,\{\,\,(\,
\frac{
-\,\ti \as (q'^2_t )}{ \,l\,+\,\gamma ( \oa_1 )}\,)^{n + 1}\,\cdot\,\Gamma
( n +1 ) \,\,+
$$
$$
\,\,\frac{1}{2}\,
( - 1 )^l  \frac{\Gamma ( \gamma( \oa_1 ) \,+\,1\,)}{ \Gamma ( \gamma ( \oa_1
)\,+\,l )\,l!}\,(\,\frac{-\,2\,\ti \as ( q'^2_t )}{l\,+\,1\,+\,\gamma(
\oa_1 )}\,)^{n
+ 1} \,\cdot\,\Gamma ( n + 1 )\,\,\}\,\,+\,\,\{ \gamma = 0 \} \,\,]\,\,=
$$
$$
 \,\,\as \,\,\Sigma_{n = 1}\,\Sigma_{l = 0}\,(\,{-\,\ti \as}\,)^n
\,R_{n,l}\,\,.
 $$
The crucial difference between \eq{SERDIFFKOALLQ} and \eq{IRSERIES} is the
alternating sign in the general term of the series of \eq{SERDIFFKOALLQ}.
Such series can give the well defined analytic function in spite of the $n!$
growth of the general term.

The physical meaning of such essential difference is very simple, because the
value of the momentum $ ( q^2_{eff})_n $ in the integral for $\Gamma (n)$ in
\eq{SERDIFFKOALLQ} turns out to be big, namely
\beq \label{QEFF}
( \,q^2_{eff}\,)_n \,\,\approx \,\,q^2_t\,\cdot\,e^{\frac{n}{\gamma (\oa_1 )}}
\eeq
and even at large value of $n$, we are still in the region of the use of
perturbative QCD.

To use this as a tool for calculation, let us consider once more
the singularities in the Borel plane.
\subsubsection{UV renormalons: singularities in the Borel plane.}
Introducing the Borel image for $\Delta \ti K_r |_{q^2_t > q'^2_t}$ as it
has been described in section 3.3.3 we see that a sum over $n$  at fixed $l$
leads to simple poles in $b$:
$$
\frac{1}{b \,\,+\,\,b_{0l}}
$$
where
\beq \label{URBO}
b_{0l} \,\,=\,\,a^{(i)}
\eeq
$$
a^{(1)}\,\,=\,\,\gamma( \oa_1 ) \,\,+\,\,l\,\,;
$$
$$
a^{(2)}\,\,=\,\,1\,\,+\,\,l\,\,;
$$
$$
a^{(3)}\,\,=\,\,\frac{1\,+\,l\,+\,\gamma ( \oa_1 )}{2}
$$
Therefore,  all singularities are located on negative real axis as shown in
Fig.3.3. It means that we can take the Borel integral without any ambiguity.
Indeed
\beq \label{BOLINTLQ}
\int^{\infty}_0 \,\,\frac{d\,b \,\,e^{-\,\frac{b}{\ti \as}}}{ b
\,\,+\,\,b_{0l}}
\,\,=\,\,-\,\,e^{\frac{b_{0l}}{\ti \as}}\,\,Ei(\,-\,\frac{b_{0l}}{\ti
\as}\,)\,\,.
\eeq
Finally the answer for $\Delta \ti K_r $ in the region $q^2_t \,>\,q'^2_t$
is:
\beq \label{DELTAKLQ}
\Delta \oa_{UV}\,\,=\,\,\Delta \ti K_r \,|_{q^2_t\,>\,q'^2_t}\,\,=\,\,
\frac{\as}{\ti \as}\,\cdot\,
\Sigma^{\infty}_{n = 0}\,\cdot
\eeq
$$
\,\{ \,\,\( e^{\,\frac{1 + n}{\ti \as}}\,\,Ei
\(-\, \frac{1 + n}{\ti \as} \)  \,\,+\,\,\frac{\ti \as}{ n + 1 }\,\,-\,\,
e^{ \frac{  n + \gamma( \oa_1 )}{\ti \as}}\,\,Ei \( \,-\, \frac{  n  +
\gamma( \oa_1 )}{\ti \as} \) \,\,+\,\,\frac{\ti \as}{ n  +  \gamma(
\oa_1 )}\,\,\) \,\,+
$$
$$
\,\,\frac{1}{2} \,( -\, 1\,)^{ n + 1 }\,\frac{\Gamma (
\gamma ( \oa_1 ) + 1 )}{ \Gamma ( \gamma ( \oa_1 ) + 1 ) \,\,n!}\,\(
-\,\frac{ 2 \ti \as}{ n + 1 + \gamma( \oa_1 ) } \,\,-\,\,e^{\frac{n + 1 +
\gamma ( \oa_1 )}{2 \,\ti \as}}\,\,Ei \(-\, \frac{n + 1 + \gamma ( \oa_1 )}{2
\,\ti \as} \)\,\,\)\,\,\} $$
\subsection{The first order correction to the intercept of the BFKL Pomeron.}
  Let us look back at  \eq{PHIONE!} and Fig. 3.1. and try to understand what
is the physical meaning of the $\Delta \ti K_r$. We can neglect at the moment
the running $\as (q'^2_t )$ in \eq{PHIONE!} for the rough estimate and
integrate  not with respect to $\oa_1$ and $\oa_2$,  but over $\gamma (
\oa_1)$ and $\gamma ( \oa_2 )$. Integrating in a such way we can see that
integral over $ q'_t $ gives us $\delta ( \gamma ( \oa_1) - \gamma (
\oa_2) ) $ and and integration over $y'$ leads to $y$ in front of the
integral. Finally, in vicinity of $\gamma ( \oa ) \,=\,\frac{1}{2} $ we
get for the $\Delta \p^{[1]}$ the answer:
\beq \label{FIRSTCOROM}
\Delta \p^{[1]} \,\,=\,\, \Delta \ti K_r ( \gamma ( \oa_1) \,=\,\frac{1}{2}
\cdot y \cdot \p ( Q^2, y )\,\,,
\eeq
where $\p (Q^2, y )$ is the solution of the BFKL equation with fixed
$\as$.

One can see from \eq{FIRSTCOROM} that $\Delta \ti K_r$ is the correction to
the value of $\ol$ in \eq{OMEGALIP}.

Fig. 3.6 shows the numerical calculations for $ \Delta
\oa_{\Sigma}\,\,= \Delta \oa_{IR}\,\,+\,\,\Delta \oa_{UV}\,\,=\,\,\Delta
\ti K_r ( \as, \gamma
= \frac{1}{2} )$ as function of $x = \frac{1}{\ti \as}$\footnote{ It should
be pointed out that the variable $x$ here, is not Bjorken variable for the
deeply inelastic scattering and we hope that the use of the same letter will
not be   misleading}.

It should be stressed that the calculation that is plotted in this figure is
completely nonperturbative one even at $x \approx 10 $. One can see this
 by eye because
the perturbative behaviour should be $\frac{1}{x^2}$ at large value of $x$.

The second remark is that the ratio $\Delta \oa_{\Sigma} / \as \chi (1/2 )
$ is smaller than 12 \% at $x > 3.5 $.
\section{The solution to the BFKL equation with  running {\bf $  \as$} .}
The numerical result of the previous section  ( $\Delta \ti K_r / \as \chi (1/2
\,\,<\,\,12\% $ ) we would like to use reducing the complete BFKL equation
with running $\as$ ( see \eq{BFKLRUN} and \eq{FINKER} ) to the equation
which is much simpler, neglecting the contribution of $\Delta \ti K_r$ in
the complete kernel of \eq{FINKER}.  This reduced equation has the form:
\beq \label{REDUCEBFKL}
\frac{ d\,\p (q^2_t, y )}{ d\,y}\,\,=\,\,\frac{N}{\pi} \,\,\int
\,\,\frac{d^2 q'_t}{\pi}\,\,\as( q'^2_t ) \,\,K ( q_t, q'_t )\,\,\p (
q'^2_t , y )\,\,\,
\eeq
where $ K (q_t, q'_t )$ is the kernel of the BFKL equation with fixed $\as$
( see  \eq{LIKERNEL} ).  The general solution of \eq{REDUCEBFKL} has been
found in the GLR paper \cite{GLR}, however we are going to discuss this
solution here for the sake of completeness, and to illustrate the connection
between this solution and anomalous dimension of \eq{LIANDI}. Using the
explicit expression for the running $\as
\,\,=\,\,\frac{2\,\pi}{b}\cdot\frac{1}{ln ( q^2/\L^2 ) }$ and the double
Mellin transform of \eq{MELLIN} we can rewrite \eq{REDUCEBFKL} for the
function $ \bar \p \,=\,\as( q^2_t) \,\p (q^2_t, y )$  in the form:
 \beq \label{REBFKLMEL}
-\,\omega \,\frac{\pa\, C (\oa, f )}{ \pa
\,f}\,\,=\,\,\frac{2\,N}{b}\,\,\chi (f ) \,\,C (\oa, f )\,\,.
\eeq
Therefore the general solution can be given in the form:
\beq \label{GENSOLRBFKL}
\bar \p ( q^2_t,y ) \,\,=\,\,\int \,\,\frac{ d\, \oa\,\,d \,\,f}{ ( 2\,\pi
\,i )^2} \ti \p (\oa ) exp \{\, \frac{2\,N}{b\,\,\oa}
\,\,\int^{\frac{1}{2}}_{f} \,\,\chi ( f' ) \,d\,f' \,\,+\,\,\oa\,y \,\,+\,(
\,f\,-\,1\,)\,r\,\} \,\,,
\eeq
where $r\,\,=\,\, ln ( q^2_t / \L^2  )$ .

One can simplify
\eq{GENSOLRBFKL} integrating over $f$ using the saddle point
approximation. The saddle point value of $f \,=\,f_S$ can be found
from the equation:
\beq \label{SADDLEF}
\as( q^2) \,\chi ( f_S ) \,\,=\,\,\oa
\eeq
The solution of \eq{SADDLEF} is the anomalous dimension of \eq{LIANDI}
($ f_S \,=\,\gamma ( \oa ) $ ). Taking the integral over $f'$ in
\eq{GENSOLRBFKL} introducing new variable $\as'$ from the equation
$$
\as' \,\,\chi (f')\,\,=\,\,\oa
$$
and integrating by parts we  derive  the answer:
\beq \label{ANDIANSWER}
\bar \p (q^2_t, y ) \,\,=\,\,\frac{1}{q^2_t} \,\,\int
\frac{d\,\oa}{2\,\pi\,i}\,\ti
\p( \oa )\,\,e^{ \oa\,y +\,\,\int_{\as( q^2_t)}\,\,\gamma(\oa, \as'
) \,\,\frac{d\,\as'}{\as'^2}}\,\sqrt{\frac{\pi\,\as (
q^2_t)}{\oa}\cdot \frac{d \gamma (\oa, \as)}{d\,\as}}
\eeq
The last factor in \eq{ANDIANSWER} comes from integration over $f$ in the
vicinity of the saddle point.  Taking into account that we can
calculate $ \frac{ d \gamma}{d \, \as}$ using only the first term in
\eq{LIANDI} we can get the following expression directly for function
$\p$:
\beq \label{ANDIANSWERFIN}
 \p (q^2_t, y ) \,\,=\,\,\frac{1}{q^2_t} \,\,\int
\frac{d\,\oa}{2\,\pi\,i}\,\ti
\p( \oa )\,\,e^{ \oa\,y +\,\,\int_{\as( q^2_t)}\,\,\gamma(\oa, \as'
) \,\,\frac{d\,\as'}{\as'^2}}
\eeq

It is easy to recognize in \eq{ANDIANSWERFIN}  the usual result of
renormalization group approach \cite{RENGROUP}.

Now let us formulate the problem which we desire  to solve. We want to
find the Green function ($ G (y - y_0, r )$ ) of the BFKL equation with
running $\as$ which satisfies the initial condition:
\beq \label{GRENINCO}
G ( y - y_0, r = r_0 ) \,\, =\,\,\delta (\, y \,-\,y_0\, )
\eeq
but in the kinematic region where the anomalous dimension $\gamma (\oa, \as )$
is close to the limited value $\gamma = \frac{1}{2}$ ( see \eq{GAMMAATOL} ).
To find such Green function it is easier to use the form of \eq{GENSOLRBFKL}
for the solution.

We would like to stress that we are going to solve the BFKL equation with
running $\as$ in usual way for the GLAP evolution equation, starting with the
$x$ distribution at
fixed initial virtuality $ q^2 = q^2_0, r = r_0 = ln ( q^2_0/\l^2 )$ and
calculating the deep inelastic structure function at larger value of $q^2$
( $q^2 > q^2_0 $). It is worthwhile mentioning that the solution of the
problem is
\beq \label{SOLFROMGREEN}
\p (q^2,y ) \,\,=\,\,\int d\, y_0 G ( y - y_0, r )\,\p_{in} ( y = y_0, r =
r_0 )
\eeq
However, the Green function for the BFKL equation is quite different from
the Green function for the GLAP one in the kinematic region, where the
anomalous dimension is close to the value of $\gamma = \frac{1}{2}$ and we
need to take into account the sum of all terms in \eq{LIANDI}.

In vicinity $ \gamma \,\ra\,\frac{1}{2}$ we can use the following expansion
for $\chi ( f )$:
\beq \label{EXPCHI}
\chi ( f ) \,\,=\,\,\chi ( \frac{1}{2} ) \,\,+\,\,14\,\zeta ( 3 ) \,(\, f -
\frac{1}{2}\, )^2\,\,=\,\,\chi ( \frac{1}{2} ) \,\cdot\,[\, 1 + \kappa \,(\,f
- \frac{1}{2} \,)^2 \,]
 \eeq
It is easy to see that such an expansion leads to \eq{GAMMAATOL} for $\gamma$.
Substituting \eq{EXPCHI} in the general solution of \eq{GENSOLRBFKL} one
can see that integral over $f$ can be taken and gives the Airy function.
The final answer for the Green function is
\beq \label{FINGREEN}
G ( y - y_0, r, r_0 ) \,\,=\,\,e^{\frac{1}{2}\,(\, r\,-\,r_0\,)} \cdot
\sqrt{\frac{r}{r_0}}\,\,\int \,\,\frac{d\,\oa}{2\,\pi\,i}\,\,\frac{ Ai \(
\,(\,\frac{\oa}{\ol\,r_0 \,\kappa}\,)^{\frac{1}{3}}\,[\, r
\,-\,\frac{\ol}{\oa}\,r_0\,] \,\)}{ Ai \(
\,(\,\frac{\oa}{\ol\,r_0 \,\kappa}\,)^{\frac{1}{3}}\,[\, r_0
\,-\,\frac{\ol}{\oa}\,r_0\,]\,\)}\,\,.
\eeq
The contour of integration with respect to $\oa$ in \eq{FINGREEN} is
located to the right of singularities of the integrant (see Fig. 4.1 ).
 Therefore, we have to know the singularities of the integrand in $\oa$ to take
the
integral over $\oa$ and the position in the $\oa$ - plane of the possible
saddle point in \eq{FINGREEN}. The Airy function is the analytic function
of its argument and the origin of singularities in \eq{FINGREEN} is the
zeros of the dominator. Within good numerical accuracy the zeros of the
Airy function can be calculated using the following equation:
\beq \label{ZEROS}
(\,\frac{\omega_{0 k}}{\ol\,r_0 \,\kappa}\,)^{\frac{1}{3}}\,[\, r_0
\,-\,\frac{\ol}{\omega_{0k}}\,r_0\,]\,\,=\,\,\(\,\frac{3}{2}\,
\(\,\frac{3}{4}\,\pi\,\,+\,\,\pi\,k\,\)\,\)^{\frac{2}{3}}\,\,.
\eeq
One can see that at very large value of $k$  $\omega_{0k}
\,\propto\,\frac{1}{k} \,\ra\,0$.  The second interesting observation is
the fact that the rightmost singularity (pole) turns out to be considerably
smaller than the value of $\ol$ ( See Table I which shows the value of six
 first $\omega_{0k}$).
\vspace{8pt}
\begin{center}
{\bf Table I}
\vspace{4pt}
\\
{$\omega_{0k}$  for the BFKL equation with  running $\as$.}
\end{center}
\vspace{4pt}
\begin{center}
\begin{tabular}{|l|l|l|l|l|l|l|}
\hline
 $\ol$ &
$\omega_{00}$ & $\omega_{01}$ & $\omega_{02}$ & $\omega_{03}$ &
$\omega_{04}$ & $\omega_{05}$ \\
\hline
 0.65 & 0.3 & 0.17 & 0.12 & 0.093 & 0.075 & 0.064 \\
\hline
 0.5 & 0.22 & 0.135 & 0.102 & 0.081 & 0.068 & 0.055 \\
\hline
 0.33 & 0.17 & 0.115 & 0.087 & 0.072 & 0.053 & 0.040\\
\hline
 0.25 & 0.14 & 0.10 & 0.078 & 0.064 & 0.056 & 0.050\\
\hline
 0.20 & 0.122 & 0.09 & 0.071 & 0.069 & 0.051 &0.046\\
\hline
\end{tabular}
\end{center}
\vspace{8pt}

Now let us discuss the saddle point in the integral of \eq{FINGREEN}. We found
such a saddle point in the kinematic region where we can use the asymptotic
expansion for the Airy function in the numerator of \eq{FINGREEN}. Indeed,
if we assume that
$$
(\,\frac{\oa}{\ol\,r_0 \,\kappa}\,)^{\frac{1}{3}}\,[\, r
\,-\,\frac{\ol}{\oa}\,r_0\,] \,\,\gg\,\,1
$$
we can use the asymptotic expansion for the Airy function which gives
\beq \label{AIRYASY}
Ai \(\,(\,\frac{\oa}{\ol\,r_0 \,\kappa}\,)^{\frac{1}{3}}\,[\, r
\,-\,\frac{\ol}{\oa}\,r_0\,] \,\)\,\,\ra\,\,exp \(\,\frac{\oa}{\ol\,r_0
\,\kappa}\,)^{\frac{1}{2}}\,[\, r
\,-\,\frac{\ol}{\oa}\,r_0\,]^{\frac{3}{2}} \,\)
\eeq
$$
\,\,\ra\,\,exp\( -
(\,\frac{\oa}{\ol\,\kappa}\,)^{\frac{1}{2}}\,\Delta r\,(\,\frac{\Delta
\oa}{\ol}\,)^{\frac{1}{2}}\,\)
$$
where $\Delta \oa\,=\,\oa\,-\,\ol$ and $\Delta r\,=\,r\,-\,r_0 $. To get
the last line in the above equation we also assumed that
\beq \label{RESTRI}
\frac{\Delta \oa}{\oa} \,\,\gg\,\,\frac{\Delta r}{r_0}
\eeq
Using \eq{AIRYASY} we can find the position of the saddle point in $\oa$,
namely
\beq \label{SADDLEOMEGA}
\oa_{S} \,\,=\,\,\ol \,\,+\,\,\frac{ (\,\Delta r\,)^2}{4\,\kappa \,\ol\,y^2}
\eeq
Therefore, the final structure of the $\oa$ - plane looks as it is shown in
Fig. 4.1. Evaluating the   integral by steepest decent method we reproduce the
solution to the BFKL equation, namely
\beq \label{DIFGREEN}
G ( y - y_0, r - r_0 )\,\,=\,\,e^{-\,\frac{1}{2}\,(\,r
\,-\,r_0\,)}\cdot\frac{1}{\sqrt{\pi \,\kappa\,\ol\,(\,y\,-\,y_0\,)}}\cdot
e^{\ol y \,-\,\frac{(\,\Delta r\,)^2}{4\,\kappa\,\ol\,(\,y\,-\,y_0\,)}}
\eeq
We can trust the above answer only in the kinematic region where
\beq \label{RREG}
(\,\Delta r\,)^2 \,\leq\, 4\,\kappa\,\ol\,(\,y\,-\,y_0\,)\,\,.
\eeq
However, we need to sutisfy also \eq{RESTRI}. Substituting \eq{RREG} in
\eq{RESTRI} we get the kinematic region where we can consider \eq{DIFGREEN}
as a solution to the problem. Namely
\beq \label{LIRE}
y\,-\,y_0 \,\,<\,\( \,\frac{r_o}{\ol \sqrt{4 \kappa \ol}}\,\)^{\frac{2}{3}}\,
\propto\,(\as ( r_0 ))^{- \frac{5}{3}}
\eeq
This value of $ y - y_0$ is still in the region of applicability of the
BFKL approach because $ \as ( q^2_0 ) ( y - y_0 ) \,\approx ( \as (
q^2_0))^{-\,\frac{2}{3} }
\,> \,1$, but both theoretically and practically it is very restricted region.

For larger value of $y = ln (1/x)$ we have to close our contour on the
poles of the dominator and we get the solution which behaves as
\beq
G ( y - y_0, r - r_0 ) \,\,\propto \,\frac{1}{x^{\oa_{00}}}
\eeq

The solution of the BFKL evolution equation with running $\as$ is given by
\eq{SOLFROMGREEN} which can be rewritten in the $\oa$ - representation in
the form:
\beq
\p( y,q^2_t  )\,\,=\,\,\int \,\,\frac{d\,\oa}{2\,\pi\,i} \cdot G ( \oa, r -
r_0 ) \cdot \p_{in} (\oa, r_0 )\,\,,
\eeq
where $\p_{in} (\oa, r_0)$ is the initial gluon distribution in the $\oa$ -
representation. We can distinguish two case with different solutions:

1. the initial distribution $\p_{in} \,\propto \,x^{- \l}$ with $\l
\,>\,\oa_{00}$. In this case we have to close the  contour on the singularities
of $\p_{in}( \oa )\, =\,\frac{\p_0}{\oa\,-\,\l}$ and the solution in the
region of small $x$  looks as follows:
$$
\p ( y = ln (1/x), q^2_t ) \,\,=\,\,\p_0 \cdot G(\oa = \l ,r - r_0 )\cdot
(\,\frac{1}{x}\,)^{\l}  $$

2. the second case is $\l \,<\,\oa_{00} $, when we have to close the contour on
the singularities of the Green function. The resulting behaviour of the
solution is closely related to the value of $\oa_{00}$ and looks as follows:
$$
\p ( y = ln (1/x), q^2_t ) \,\,=\,\,\p_{in} ( \oa = \oa_{00} )
 \cdot \{\,G(\oa
,r - r_0 )\cdot (\, \oa - \oa_{00} \,)\,\}\,|_{\oa = \oa_{00}}\cdot
(\,\frac{1}{x}\,)^{\oa_{00}}\,\,=  $$
$$
\p_{in} ( \oa = \oa_{00} ) \cdot\,e^{ \frac{1}{2} \,(\,r\,-\,r_0\,)}\cdot
e^{-\,\frac{2}{3} \,\(
\frac{\oa_{00}}{\ol\,r_0\,\kappa}\)^{\frac{1}{2}}\,\,\( -\,\frac{\ol}{\oa_{00}}
\,r_0\,-\,r \)^{\frac{3}{2}}}\,\cdot \( \,\frac{1}{x}\,\)^{\oa_{00}}\,\,.
$$
The above formula solves the problem.
\section{Conclusion.}
In this paper we attempted to discuss  the BFKL equation with
the running $\as$ in a systematic way. Our results look as follows:

1. We found that  this equation has the form of eq. ( 50 ) with the kernel of
eq. ( 51 ).
The weakness of the argumentation stems from the fact that we assumed the
bootstrap equation to reconstruct the form of the kernel. It is not clear
how general the bootstrap property of the BFKL equation is. To check this we
have to calculate the sets of the Feynman diagrams shown in Figs. 2.9 and
2.11. These calculations are now  in progress  and will be published elsewhere
soon.

2. The uncertainties from the contribution of the infrared renormalons in
the BFKL equation with running $\as$ were estimated and it was shown that
they are smaller that the nonperturbative contribution describing the
shadowing correction in the deeply inelastic scattering.

3. It was shown that the infrared renormalons give rise to the corrections
of the order of $\frac{1}{\sqrt{Q^2}}$ to the gluon structure function in
the region of small $x$. The physical origin of such sort  corrections as
well as their  selfconsistent nonperturbative description is still unclear  and
has
to be clarified in future.

4. The analytic function summing the infrared and ultraviolet renormalons in
the BFKL equation with running $\as$ was suggested, and numerical estimates
were given which led to simplification of the answer in the deeply inelastic
kinematic region.

5. The reduced evolution equation with the running $\as$ based on the
numerical estimates of the nonperturbative contribution to the kernel of the
BFKL equation with running $\as$ was proposed and  solved. The
result of the solution shows much slower behaviour of the gluon structure
function in the region of small $x$,  than it was predicted in the original
version of the BFKL equation with fixed $\as$.

6. The legitimate theoretical region $\as \ln ( 1/ x) \,<\,( \as( q^2_0)
)^{\frac{2}{3}}$ for the BFKL equation with fixed $\as$ was found.

We hope that this paper will be useful in understanding what kind of
 nonperturbative phenomena has been taking into account in the BFKL equation
and
for   more elegant and comprehensive theoretical description of nonperturbative
QCD in the region of low $x$ ( high energies).

{\bf Acknowlegements:} We would like to thank M. Braun for valuable
discussions on the subject  as well as for  sending us  his
paper \cite{BRAUN} before publication. We are grateful to all participants of
high energy seminar at Tel Aviv University and at the  LAFEX,CBPF and
especially to A.
Gotsman for useful comments
on the paper. We acknowledge the financial  support by Mortimer and Raymond
Sackler Institute of Advanced Studies  and by  the CNPq.
 
\newpage
\section*{Figure Captions}
\begin{tabular}{l l}
{\bf Fig. 2.1\,\,:} &\,\,\,\,The Born Approximation of perturbative QCD for
quark - quark\\
 &\,\,\,\, scattering.\\
  & \\
{\bf Fig. 2.2\,\,:}&\,\,\,\,The next to Born Approximation of perturbative
QCD for quark - quark\\
 &\,\,\,\, scattering:\,\,  emission of one extra gluon.\\
 & \\
{\bf Fig. 2.3\,\,:}&\,\,\,\, The next to Born Approximation of perturbative
QCD for quark - quark\\
 &\,\,\,\, scattering:\,\,$\as^2$ correction to elastic amplitude.\\
 & \\
{\bf Fig. 2.4\,\,:}&\,\,\,\,The next to Born Approximation of perturbative
QCD for quark - quark\\
& \,\,\,\,scattering:\,\, gauge invariance trick.\\
 & \\
{\bf Fig. 2.5\,\,:}&\,\,\,\, The next to Born Approximation of perturbative
QCD for
quark - quark\\
 &\,\,\,\,  scattering:\,\,the resulting answer for emission of one extra
gluon.\\
 & \\
{\bf Fig. 2.6\,\,:}& \,\,\,\,The relation between colour coefficients.\\
 & \\
{\bf Fig. 2.7\,\,:}&\,\,\,\,The BFKL equation.\\
      & \\
{\bf Fig. 2.8\,\,:}& \,\,\,\,The Born Approximation of perturbative QCD for
quark - quark\\
 &\,\,\,\, scattering  for a running coupling constant.\\
 & \\
{\bf Fig. 2.9\,\,:}& \,\,\,\,Insertion of fermion bubles in the amplitude of
emission of one extra gluon.\\
 & \\
{\bf Fig. 2.10\,\,:}&\,\,\,\, Corrections to the reggeization of the gluon
due to running $\as$.\\
 & \\
{\bf Fig. 2.11\,\,:}& \,\,\,\, Emission of one quark - antiquark pair.\\
 & \\
{\bf Fig. 3.1\,\,:}& \,\,\,\,The first correction to the BFKL equation due to
running $\as$.\\
 & \\
{\bf Fig. 3.2\,\,:}&\,\,\,\, The shadowing correction to the deep inelastic
scattering.\\
 & \\
{\bf Fig. 3.3\,\,:}& \,\,\,\,The structure of singularities in the  Borel
plane.\\
 & \\
{\bf Fig. 3.5\,\,:}& \,\,\,\,$\Delta \oa_{IR}$ versus $\gamma$ at fixed
$\as$\,=\,0.25.\\
 & \\
{\bf Fig. 3.5\,\,:}& \,\,\,\,$\Delta \oa_{IR}$   versus
$x\,\,=\,\,\frac{1}{\ti \as}$ at fixed $ \gamma\,\,=\,\, \frac{1}{2}$.\\
  &  \\
{\bf Fig. 3.6\,\,:}& \,\,\,\,$\Delta \oa_{\Sigma} $ versus
$x\,\,=\,\,\frac{1}{\ti \as}$ at fixed $ \gamma\,\,=\,\, \frac{1}{2}$.\\
 & \\
{\bf Fig. 4.1\,\,:}&\,\,\,\, The structure of singularities in the $\oa$ -
plane for the solution of\\
& \,\,\,\, the reduced BFKL equation with running $\as$.\\
\end{tabular}

\end{document}